\documentclass{emulateapj}

\begin{document}
\title{Far-ultraviolet Observations of the Taurus-Perseus-Auriga Complex}

\author{Tae-Ho Lim\altaffilmark{1}}
\affil{Korea Advanced Institute of Science and Technology (KAIST)}
\email{thlim@kaist.ac.kr}

\author{Kyoung-Wook Min\altaffilmark{1}}
\affil{Korea Advanced Institute of Science and Technology (KAIST)}

\and

\author{Kwang-Il Seon\altaffilmark{2}}
\affil{Korea Astronomy and Space Science Institute (KASI)}

\altaffiltext{1}{Korea Advanced Institute of Science and Technology (KAIST), 373-1
Guseong-dong, Yuseong-gu, Daejeon, Korea 305-701, Republic of Korea}

\altaffiltext{2}{Korea Astronomy and Space Science Institute (KASI), 61-1 Hwaam-dong,
Yuseong-gu, Daejeon, Korea 305-348, Republic of Korea}

\begin{abstract}
We have constructed a far-ultraviolet (FUV) continuum map of the Taurus-Auriga-Perseus complex, one of the largest local associations of dark clouds, by merging the two data sets of GALEX and FIMS, which made observations at similar wavelengths. The FUV intensity varies significantly across the whole region, but the diffuse FUV continuum is dominated by dust scattering of stellar photons. A diffuse FUV background of $\sim$1000 CU is observed, part of which may be attributable to the scattered photons of foreground FUV light, located in front of the thick clouds. The fluorescent emission of molecular hydrogen constitutes $\sim$10 \% of the total FUV intensity throughout the region, generally proportional to the local continuum level. We have developed a Monte Carlo radiative transfer code and applied it to the present clouds complex to obtain the optical properties of dust grains and the geometrical structures of the clouds. The albedo and the phase function asymmetry factor were estimated to be $0.42^{+0.05}_{-0.05}$, and $0.47^{+0.11}_{-0.27}$, respectively, in accordance with theoretical estimations as well as recent observations. The distance and thickness of the four prominent clouds in this complex were estimated using a single slab model applied individually to each cloud. The results obtained were in good agreement with those from other observations in the case of the Taurus cloud, as its geometrical structure is rather simple. For other clouds, which were observed to have multiple components, the results gave distances and thicknesses encompassing all components of each cloud. The distance and thickness estimations were not crucially sensitive to the exact values of the albedo and the phase function asymmetry factor, while the locations of the bright field stars relative to the clouds as initial photon sources seem to be the most important factor in the process of fitting.
\end{abstract}


\section{Introduction}
The immense complex of the Taurus, Perseus, and Auriga (TPA) clouds is known to be one of the largest local associations of dark nebula. According to \citet{ung87}, the total masses of the whole complex is $\sim 2 \times 10^5$ solar mass. The TPA complex contains star-forming cloudlets (TMC1, TMC2, NGC 1333, and IC 348, for example), a bright emission nebula (NGC 1499, also known as California Nebula), and stellar associations (Per OB2, for example). Low-mass star formation processes are believed to be taking place in the eastern section of the complex (Taurus and Auriga), where mostly low-mass T Tauri type stars are observed, while both low- and high-mass stars are being formed in the western section of the Perseus complex (the Per OB2 association, the open cluster NGC 1333, and so on). The whole complex may be physically associated with the Per OB2 stellar association, centered at (l, b)=(160.5, -14.7) with a diameter of $\sim20^{\circ}$, which is the second closest OB association to the Sun with a distance of $\sim$300 pc on average \citep{bel02a}. Recent supernova explosions in the Per OB2 association are believed to have created an expanding HI supershell that is moving into the surrounding interstellar medium \citep{san74, har97}.

The TPA complex may consist of multiple layers of molecular gas located at various distances. The Taurus cloud is believed to spread between 100 pc and 200 pc \citep{ung87}, with a large part of it blocked by the cloud itself \citep{eli78,ken94}. The most distant part of the Taurus molecular cloud complex may be interacting with a supershell blown by the Per OB2 association \citep{ola87}. \citet{cer90} argued that the cloud in the region of NGC 1333, one of the star forming regions located in and around the Perseus cloud, consists of two layers, and estimated their distances to be $\sim$170 pc and $\sim$230 pc based on an extinction analysis. The Auriga dark cloud, located at a distance similar to that of the Taurus cloud, is usually regarded as part of the Taurus-Auriga complex, and was suggested to overlap with the California cloud at its farthest side \citep{ung87}. The California cloud seems to be the most distant cloud in this region: \citet{lad09} argued that the cloud is located at $\sim$450 pc from the Sun.

The interstellar radiation field (ISRF), composed of radiations originating directly from stellar sources as well as the diffuse background, plays an important role in the physical and chemical processes of the interstellar medium (ISM). For example, far-ultraviolet (FUV) ISRF photons may ionize atoms, dissociate molecules, or heat gases by ejecting electrons from dust grains or by directly exciting atoms and molecules. The ISRF itself is also affected by interactions with the ISM. In fact, most of the diffuse FUV background radiation is believed to originate from the scattering of starlight by interstellar dust \citep{bow91,seo11}. The effect of dust scattering is well observed in reflection nebulae and dark clouds as well as in the diffuse galactic light (DGL). Hence, the observations of these objects have been used not only to trace interstellar dust, but also to obtain the optical properties of dust itself, such as the albedo (a) and the phase function asymmetry factor (g). In the case of the DGL, dust scattering was observed to be moderately reflective in the forward direction \citep{lil76,jou83,hur91,mur91,mur94,mur95,wit97}, with the estimated values of a and g in general agreement with those of the theoretical model for carbonaceous-silicate grains, i.e., 0.40-0.60 and 0.55-0.65, respectively, at UV wavelengths \citep{wei01}. For reflection nebulae \citep{mur93,wit92,gor94,gib03} and individual clouds \citep{wit90,mat70,hur94}, however, the results were varied and dependent on the individual targets, probably due to uncertainties in the scattering geometry of the target clouds as well as the star systems that were the dominant sources of photons \citep{dra03,bow91}. \citet{cal95} determined a and g to be 0.7-0.8 and $\sim$0.75 around 1600 \AA, respectively, for dust toward IC 43, while \citet{gib03} estimated a and g to be $\sim$0.22 and $\sim$0.74, respectively, for wavelengths below 2200 \AA $ $ for the Pleiades nebula. For the Ophiuchus cloud, \citet{suj05} found a and g to be $\sim$0.40 and $\sim$0.55, respectively, at $\sim$1100 \AA, while \citet{lee08} obtained $\sim$0.36 and $\sim$0.52 for a and g, respectively, for 1370 \AA--1670 \AA.

\begin{figure*}
	\begin{center}
	\begin{tabular}{cc}
  		\includegraphics[width=8cm]{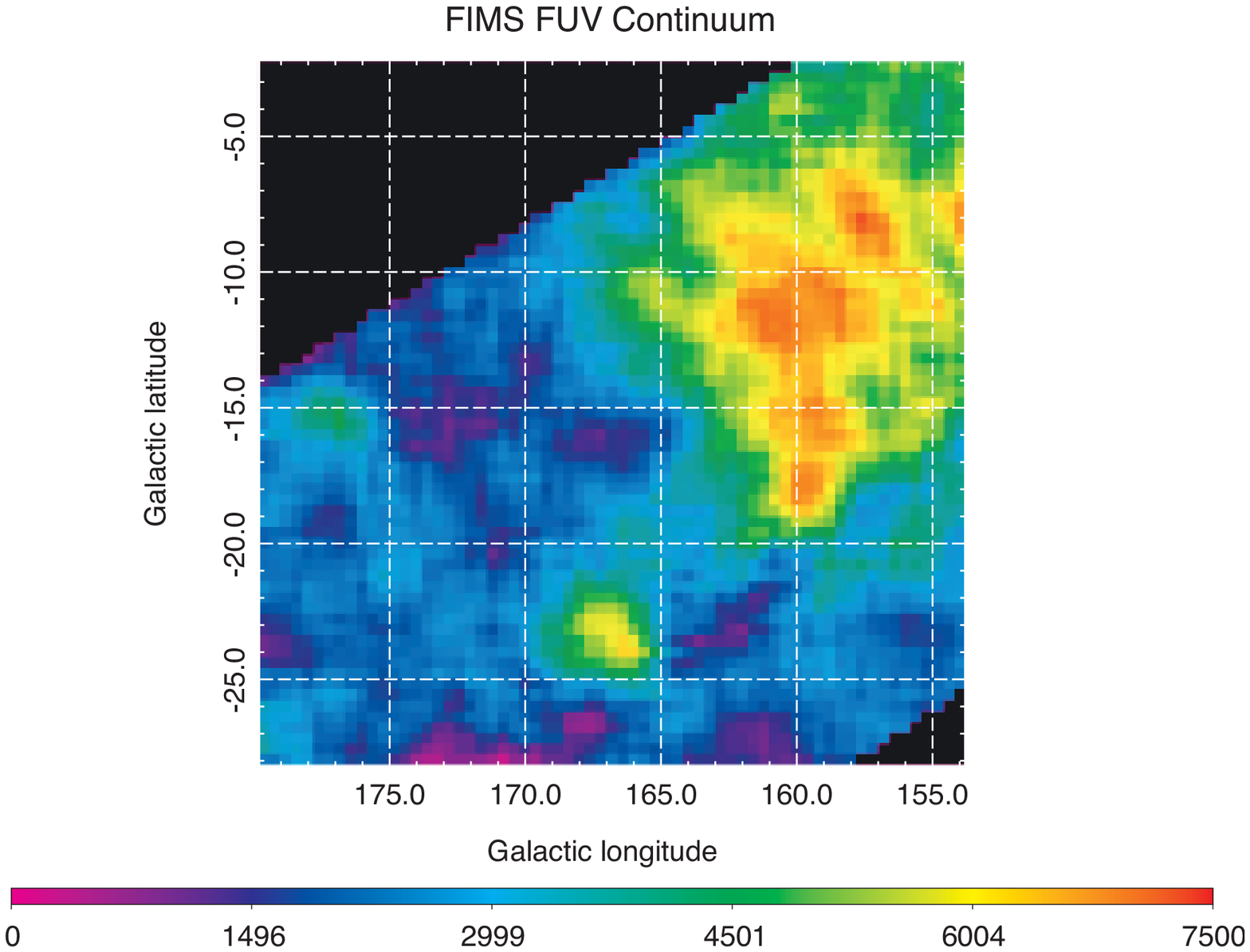} &
  		\includegraphics[width=8cm]{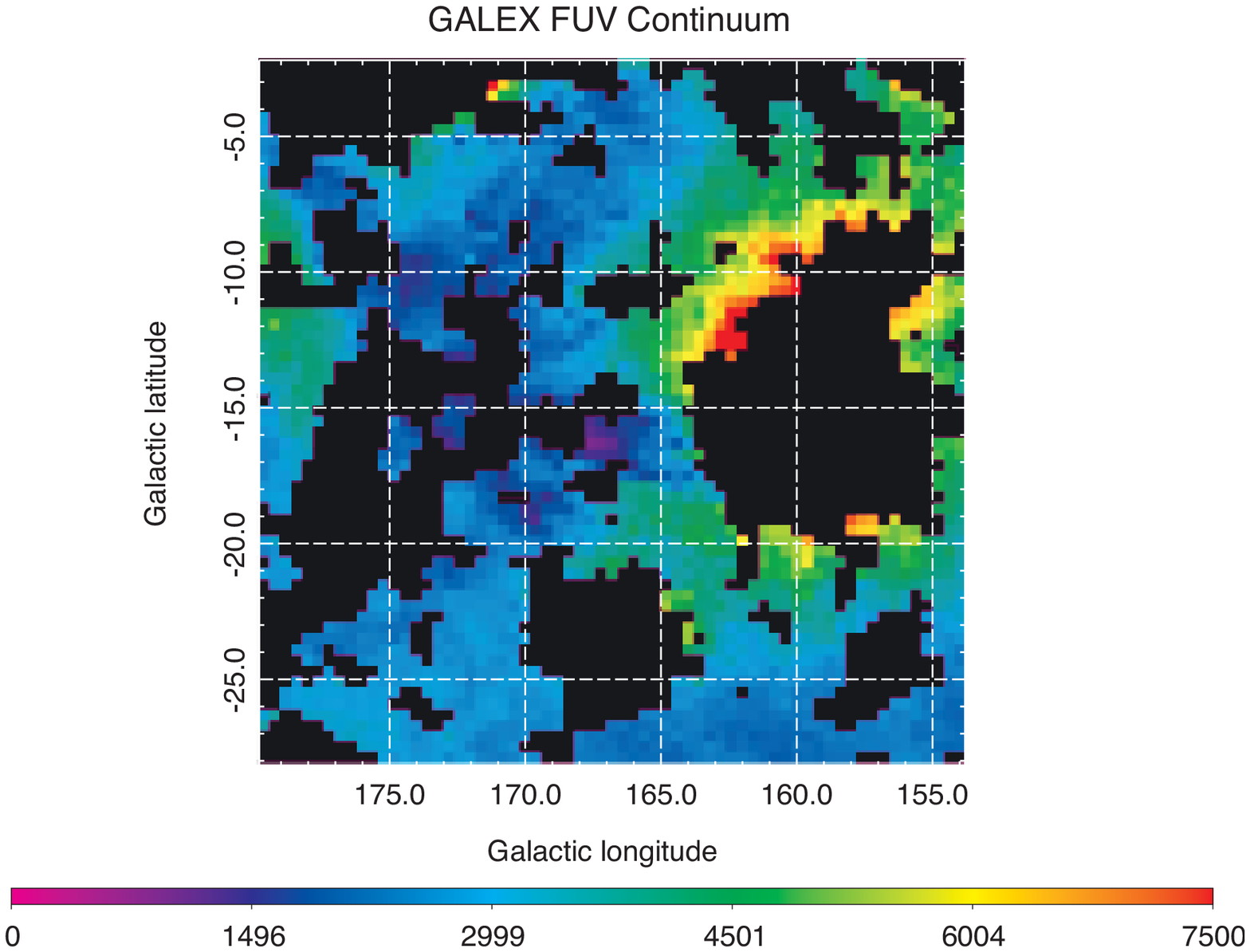}
	\end{tabular}
	\end{center}
	\caption{The FIMS continuum image (a) and the GALEX FUV image (b). \label{fig:fng}}
\end{figure*}

The main goal in this line of research has been the estimation of the optical properties of dust grains. The geometry of the clouds and the distribution of stellar photon sources involved, however, were also assumed to influence the intensity of light scattered by dust in the clouds, thereby providing constraints on the geometrical structures of the clouds themselves. For example, \citet{suj05} assumed two extended sheet-like structures for their study on the Ophiuchus cloud. \citet{sha06} also used the morphology of a plane-parallel dust sheet in their scattering simulation of the Orion nebula. Although these assumed clouds geometries were highly idealistic, the estimated distances were more or less reasonable when compared to results based on other observations. In this paper, we examine the results of the FUV observations made for the TPA region and study the clouds geometries, though the results might still be very simple. We perform Monte Carlo simulations of dust scattering and compare the results with the observed diffuse FUV emission map. In fact, it was revealed in the study of the Taurus region that the continuum emission is closely related to dust extinction \citep{lee06}. There have been several reports on the structures of the clouds in this region, mostly based on star counting methods, extinction studies, and comparisons with the distances of the associated stars and nebulae. We show that the present method of dust scattering simulations gives consistent estimates regarding the clouds' geometries in this region. We also demonstrate that the average albedo value obtained for this region is in good agreement with a theoretical estimation.

\section{Data Reduction}
We employed two data sets for the present study. The first is from observations made by the FUV Imaging Spectrograph (FIMS) flown aboard the Korean microsatellite STSAT-1, which was launched on September 27, 2003 and operated for a year and a half. The second consists of FUV archival data of the Galaxy Evolution Explorer (GALEX). FIMS is an imaging spectrograph optimized for the observation of diffuse emissions with two wavelength channels, a short wavelength channel covering $\lambda\lambda$ 900--1150 \AA, and a long wavelength channel covering $\lambda\lambda$ 1350--1700 \AA. Both channels have a resolving power of $\lambda/\Delta\lambda \sim$ 550. We analyzed only the long wavelength channel data in the present study since the short wavelength channel was contaminated by many airglow lines. The field of view of the long wavelength channel is $7.4^\circ \times 4.3'$, with a 5--10$'$ imaging resolution along the slit, and the spectral half-energy line width, averaged over the angular field, is 3.2 \AA. The instrument, its in-orbit performance, and the data analysis procedures are described in Edelstein et al. (2006a, 2006b).

The FIMS data set that covers the TPA area of 154$^\circ$ -- 180$^\circ$ in Galactic longitude and -28$^\circ$ -- -2$^\circ$ in Galactic latitude is composed of 38 observations made during the sky survey toward this region. The total exposure time is around 3.6 s per pixel, on average. Only the data from 1360 \AA $ $ to 1680 \AA, excluding the intense O {\scriptsize I} airglow line at 1356 \AA, was used for the present analysis. The FIMS continuum image of the diffuse emission was constructed by spatially binning the data with a square bin of 0.4$^\circ$ x 0.4$^\circ$ size, after removing pixels that had more than 9,000 Continuum Unit (photons s$^{-1}$ cm$^{-1}$ sr$^{-1}$ \AA$^{-1}$, hereafter CU) as these were suspected to be contaminated by bright stars. The 9000 CU criterion was chosen as the data of GALEX, which intentionally avoided the observation of bright stars, did not exceed 9000 CU in this region.
Figure 1(a) shows the resulting FIMS image, where the regions of missing data are shown in black. While the map shows the expected general features of the TPA region, including the dark Taurus cloud and the bright Per OB2 association, we can also see that the bright regions of the Per OB2 association are very extended across the scanning direction. In fact, this extended appearance is due to instrumental scattering along the slit. Furthermore, some areas of the FIMS image were made with extremely short exposures. Hence, we merged the GALEX FUV data with the FIMS data to remedy these shortcomings, as the wavelength range of the GALEX data is similar to that of the FIMS, i.e., 1350 -- 1750 \AA.

From the GALEX data release website, we extracted the All Sky Surveys imaging data (AIS of GR4/GR5 Data Release) of the TPA region, and plotted the image with 0.4$^\circ$ x 0.4$^\circ$ bins in Figure 1(b). We compared the intensities of the two observations pixel by pixel to check for possible biases that might exist between them. We limited our comparison to only the regions in which the FIMS intensity was less than 6000 CU, to exclude the bright pixels affected by the instrumental scattering and the pixels where FIMS exposure time was greater than 2.5 s. In Figure 2, we show the resulting scatter plot, in which we can see a remarkable correlation between the two data sets. The linear fitting between the two intensities is given by $I_{GALEX} = 1.01 \times I_{FIMS} + 185.50$ in CU. The standard deviation of the data about the linear fit is $\sim$ 550 CU. Since we believe GALEX data are more reliable than FIMS data in the region where the GALEX data are available because of instrumental scattering and short exposure times of FIMS, we used GALEX data as the basis of the merged map and filled the regions where GALEX did not observe with the FIMS data scaled according to the above linear fitting function. The resulting merged map is shown in Figure 3(a).

\begin{figure}
	\begin{center}
  		\includegraphics[width=9cm]{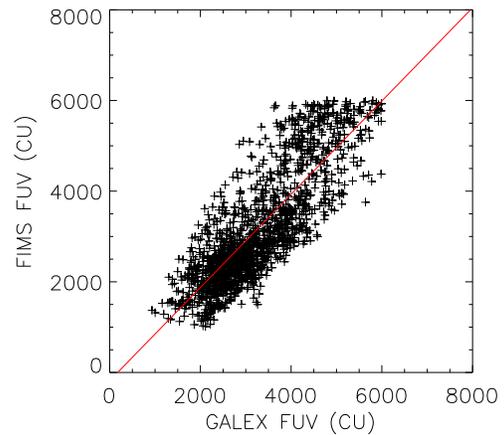}\\
	\end{center}
	\caption{A pixel-to-pixel scatter plot for the intensities of the TPA region observed by FIMS and GALEX showing a remarkable agreement. The red solid line is a linear fit of the data points. \label{fig:fng_p2p}}
\end{figure}

\section{Results}
\subsection{Examination of the FUV continuum map}

\begin{figure*}
	\begin{center}
	\begin{tabular}{cc}
  		\includegraphics[width=8cm]{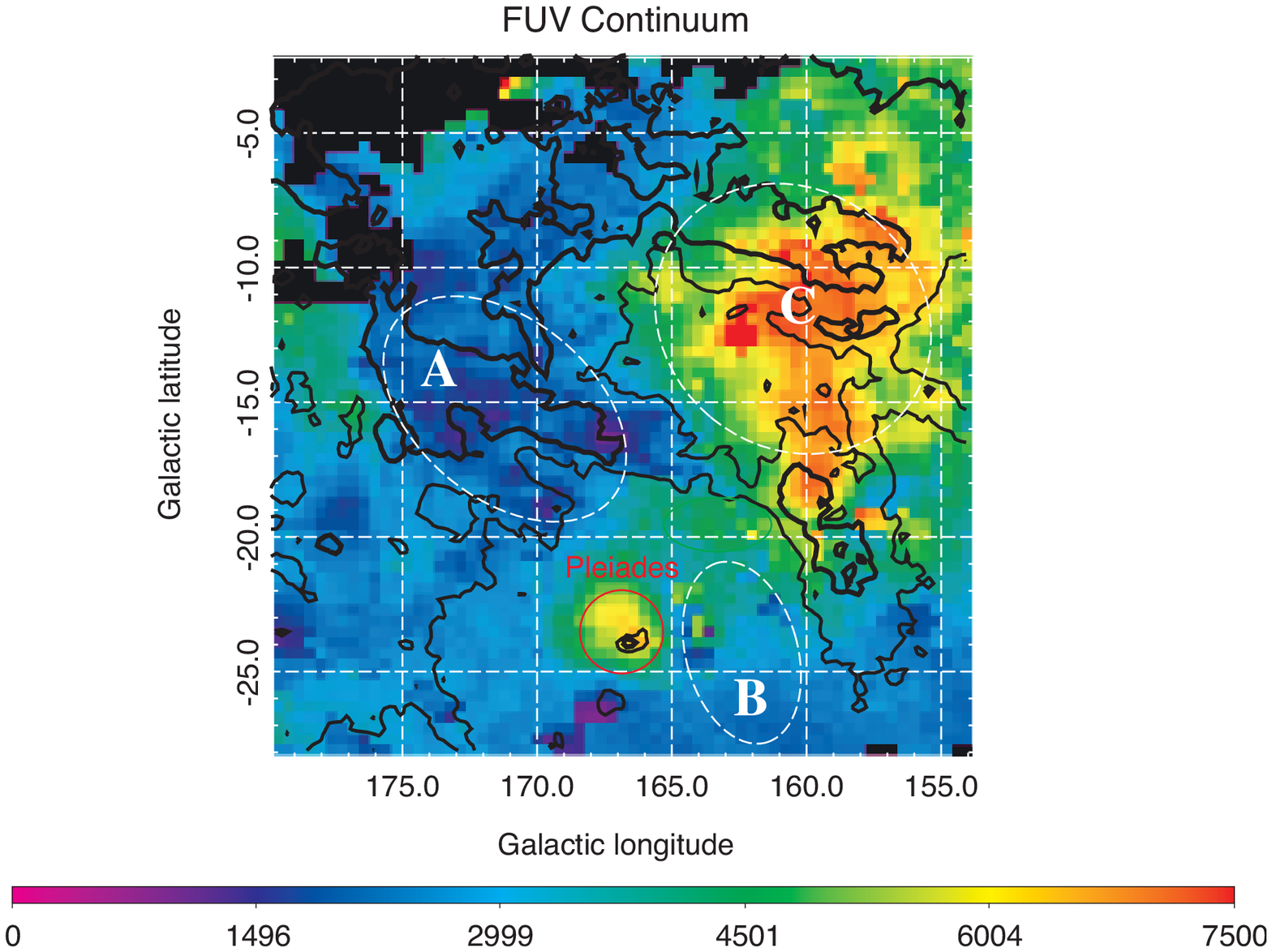} & 
		\includegraphics[width=8cm]{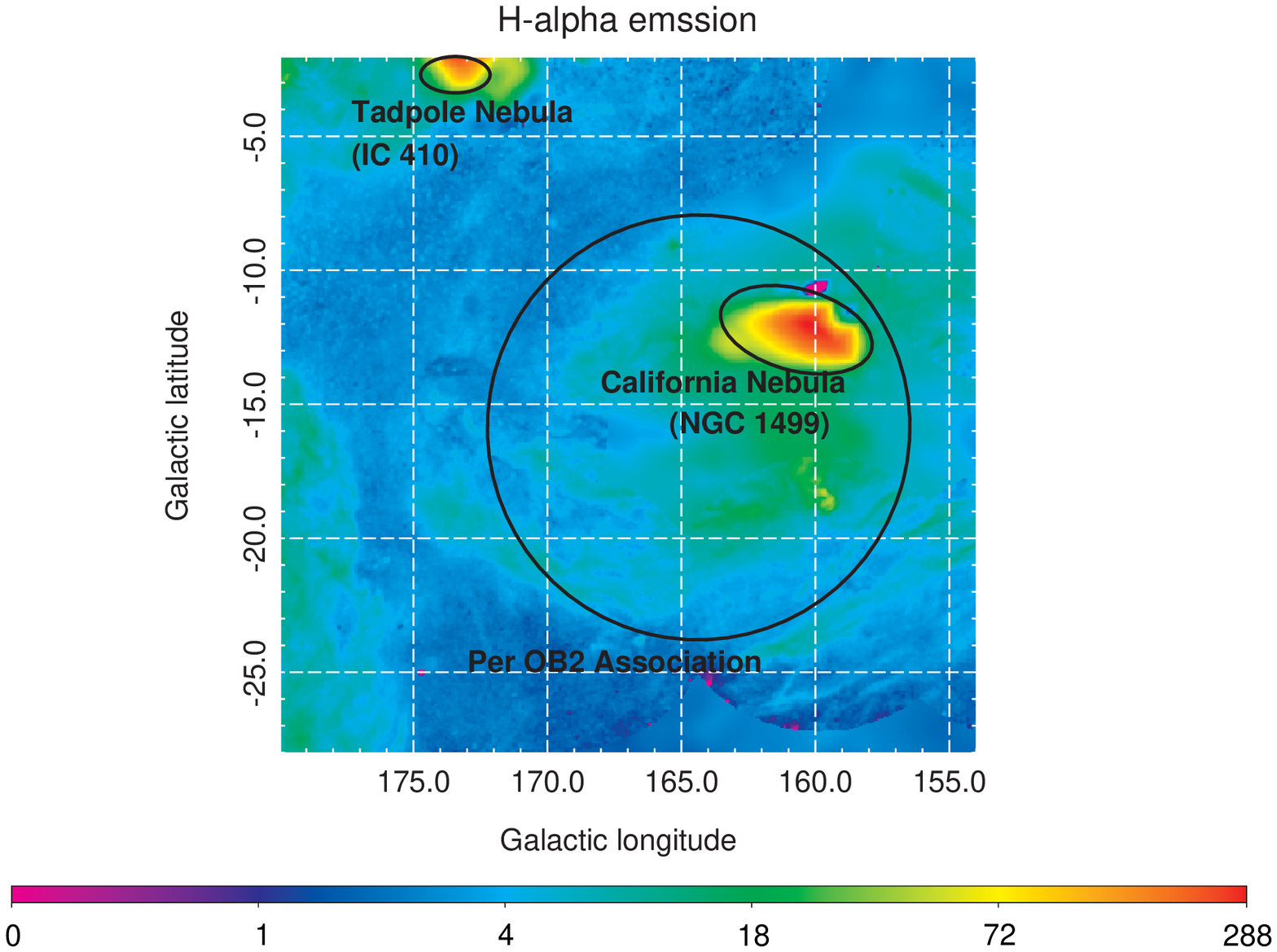} \\
  		\includegraphics[width=8cm]{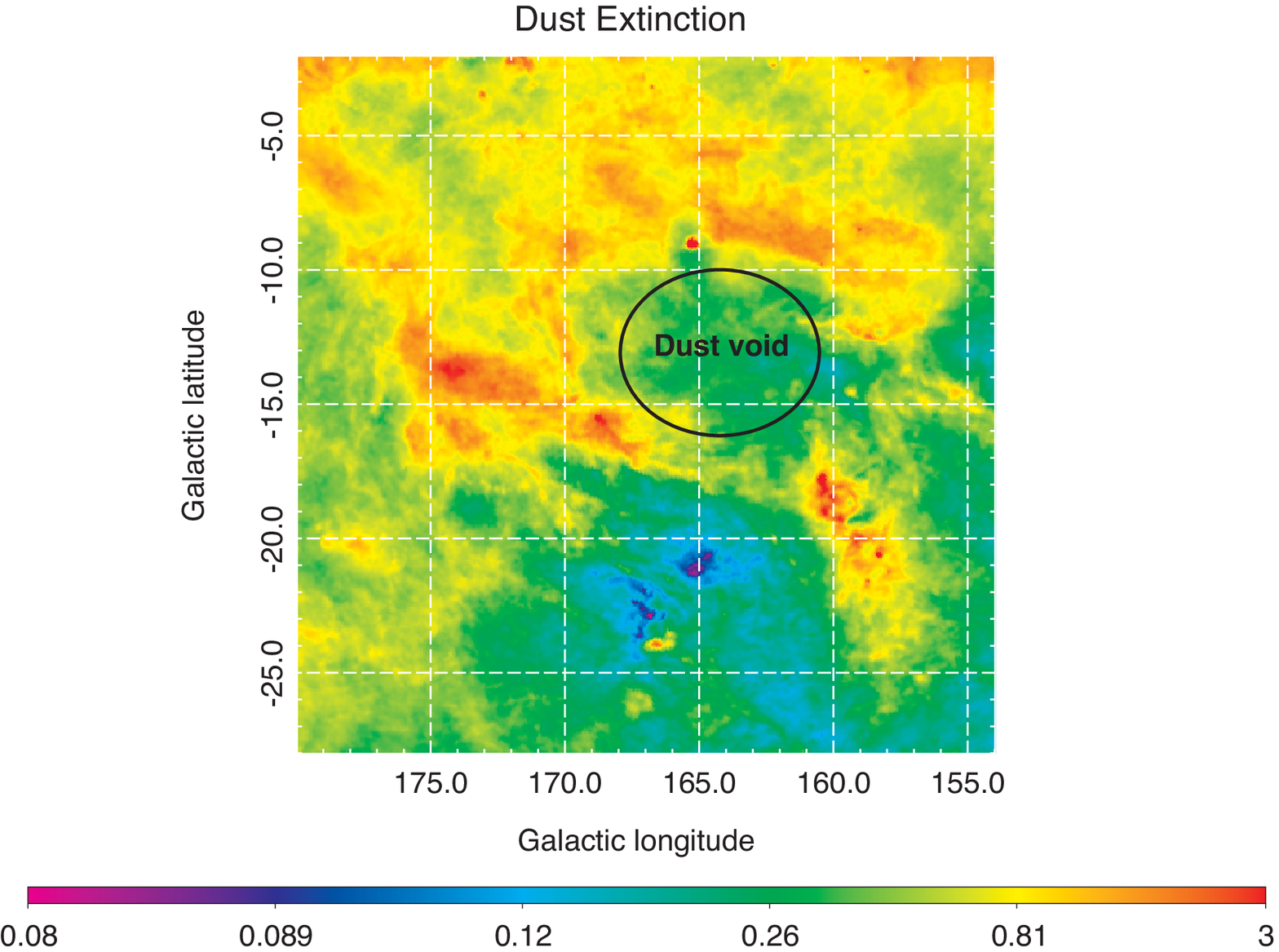} & 
		\includegraphics[width=8cm]{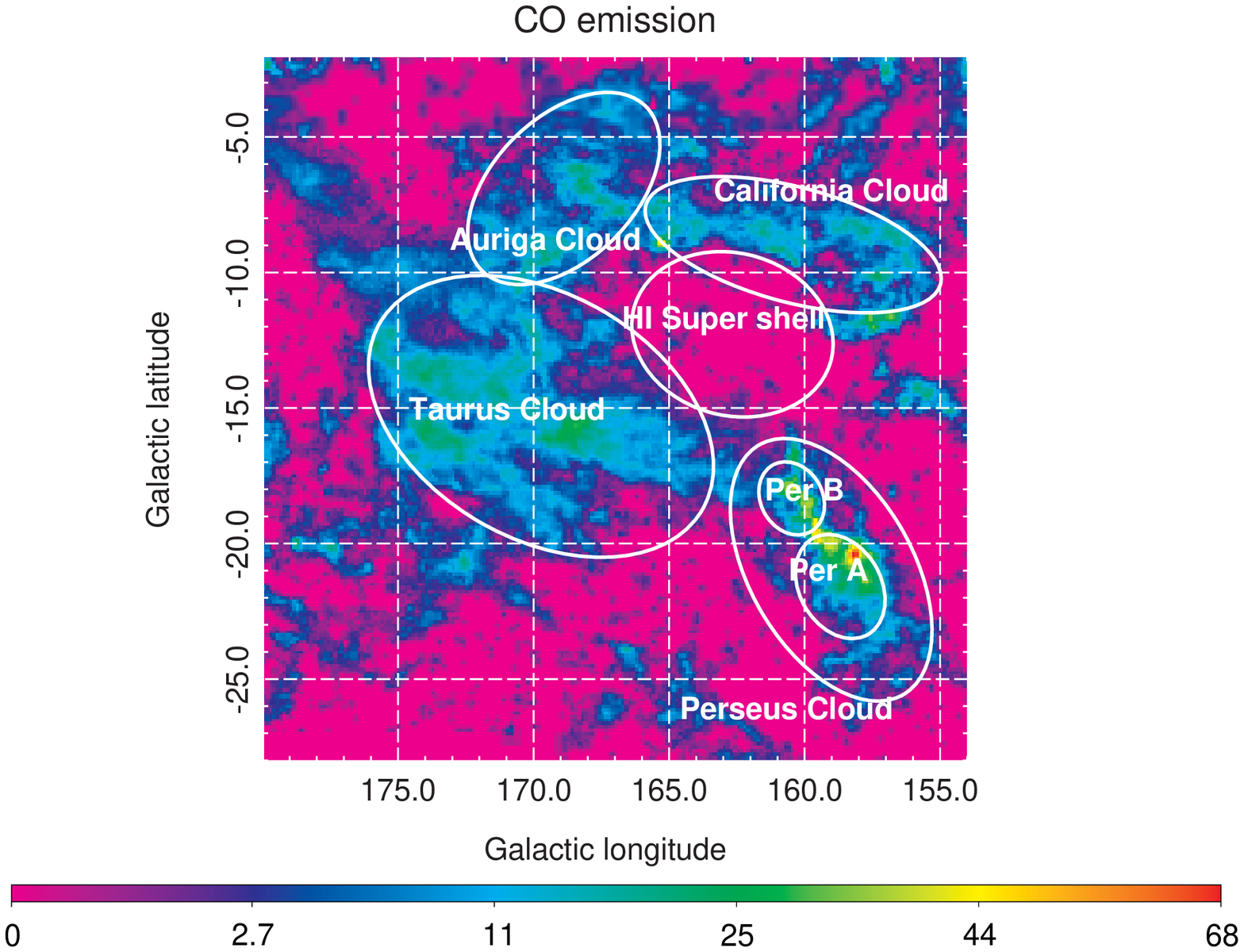} 
	\end{tabular}
	\end{center}
	\caption{(a) FUV intensity (in CU) overlaid by the dust extinction contours, with thicker contours representing higher dust extinction levels, (b) H$\alpha$ emission map shown in Rayleighs, (c) SFD dust extinction map, E(B-V), and (d) CO emission map, shown in K km/s. The locations A, B, and C in panel (a) are those with distinct physical characteristics and discussed in the text.  \label{fig:multiwave}}
\end{figure*}

Figure \ref{fig:multiwave} shows images of the TPA region constructed at different wavelengths: (a) the FUV map obtained by merging the FIMS and GALEX data, (b) H$\alpha$ emission map \citep{fin03}, (c) Schlegel, Finkbeiner \& Davis (SFD) Dust Survey map \citep{sch98}, and (d) the CO emission map \citep{dam01}. The FUV continuum map of Figure \ref{fig:multiwave}(a) shows bright enhancement in the regions of the Per OB2 association and the Pleiades cluster, as well as dark regions coincident with the Taurus and Auriga clouds. Hence, the FUV flux is bright on the low-longitude half where H$\alpha$ emission is also bright, while FUV flux seems to be heavily absorbed by the dense cloud of the Taurus-Auriga complex in the high-longitude half, where the dust extinction level is high. The existence of a dense cloud, however, does not necessarily imply that FUV intensity will be low there. For example, the California cloud near the Per OB2 association is rich in dust but FUV intensity is still bright there, a feature related to their respective distances, to be discussed later. In the dust extinction and CO emission maps, shown in Figure \ref{fig:multiwave}(c) and Figure \ref{fig:multiwave}(d), respectively, a common void or cavity is seen surrounded by the dark clouds of the TPA complex. The void is coincident with the brightest region in the H$\alpha$ emission map of Figure \ref{fig:multiwave}(b) in the vicinity of the center of the Per OB2 association. Though not noticeable here, a dust emission ring is seen around the cavity according to sub-millimeter and far-infrared maps \citep{wat05}. In fact, the void region was identified to be an HI supershell with a diameter of $\sim$100 pc, which is believed to have been formed by the interaction of the Per OB2 association with the ambient interstellar medium \citep{san74,hei84}. The bright spot south of the Taurus cloud is the reflection nebula around the Pleiades cluster.

\begin{figure}
	\begin{center}
  		\includegraphics[width=6.5cm]{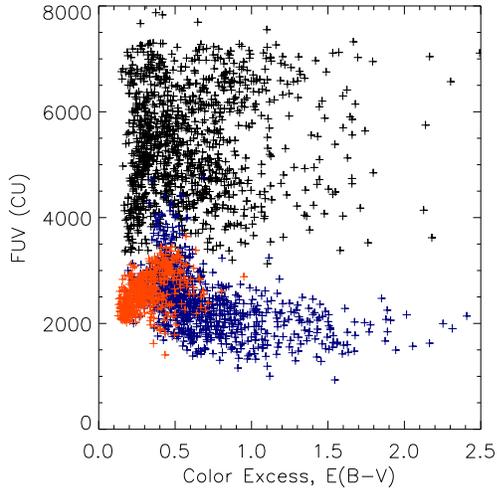}\\
	\end{center}
	\caption{A scatter plot of the FUV intensity against dust extinction: the blue, orange, and black colors correspond to the regions A, B, and C in Figure \ref{fig:multiwave}(a), respectively.  \label{fig:fuvscatt}}
\end{figure}

We have selected three distinct subregions as shown in Figure \ref{fig:multiwave}(a) and made a pixel-to-pixel plot of the FUV intensity against dust extinction, with different colors according to the subregions. Subregion A includes the Taurus cloud which is optically thick with low FUV intensity, and subregion B is optically thin with a little higher FUV intensity. Subregion C is a region of high FUV intensity that includes both optically thin and thick areas. The result is shown in Figure \ref{fig:fuvscatt}. First, we note that region A, which includes the thick Taurus cloud, indeed shows a general anti-correlation between the FUV intensity and the dust extinction level, implying that FUV is heavily absorbed by the dense Taurus cloud. On the other hand, the result for region B shows a positive correlation, in agreement with the general notion that FUV intensity increases with dust extinction level in the optically thin region \citep{hur94,luh96}. Indeed, most of the data points in region B correspond to small values of color excess, below $\sim$0.5. On the other hand, region C shows severe fluctuations, obviously due to the non-uniform strong radiation fields of the Per OB2 association over a wide range of the color excess values. Another interesting feature seen in Figure \ref{fig:fuvscatt} is that FUV intensity does not decrease below 1000 CU even at a very high extinction level, part of which we may attribute to the scattered photons of the foreground FUV light, located in front of the thick clouds \citep{lee06}. \citet{hur91} also observed a similar diffuse background for the Galactic latitudes $10^\circ < |b| < 20^\circ$.

\subsection{Monte Carlo simulation of dust scattering}

As shown in Figure \ref{fig:fuvscatt}, the scatter plot of FUV intensity against the dust extinction level shows complicated behaviors that depend on local extinction levels, fluctuations of local radiation fields as well as mixing of the foreground and background FUV photons. On the other hand, such complicated behaviors might also make it possible to extract information regarding the structure of the clouds, especially their relative locations, by comparing the observed FUV intensities with those of dust-scattering simulations that incorporate local photon sources and an assumed distribution of clouds. In fact, similar methods have been previously applied to several cloud systems \citep{gor01,sha06,lee08,suj05,suj07}.

\begin{figure*}
	\begin{center}
	\begin{tabular}{cc}
  		\includegraphics[width=8cm]{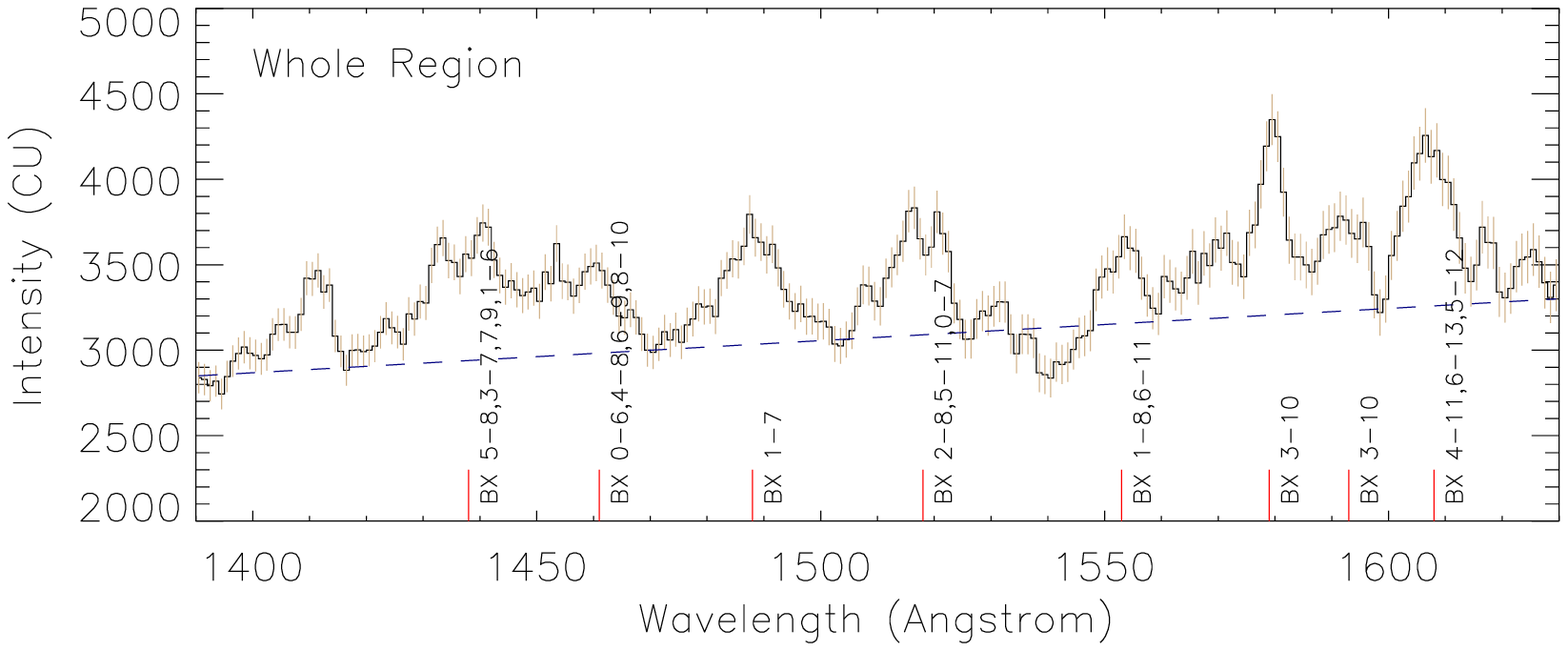} &
  		\includegraphics[width=8cm]{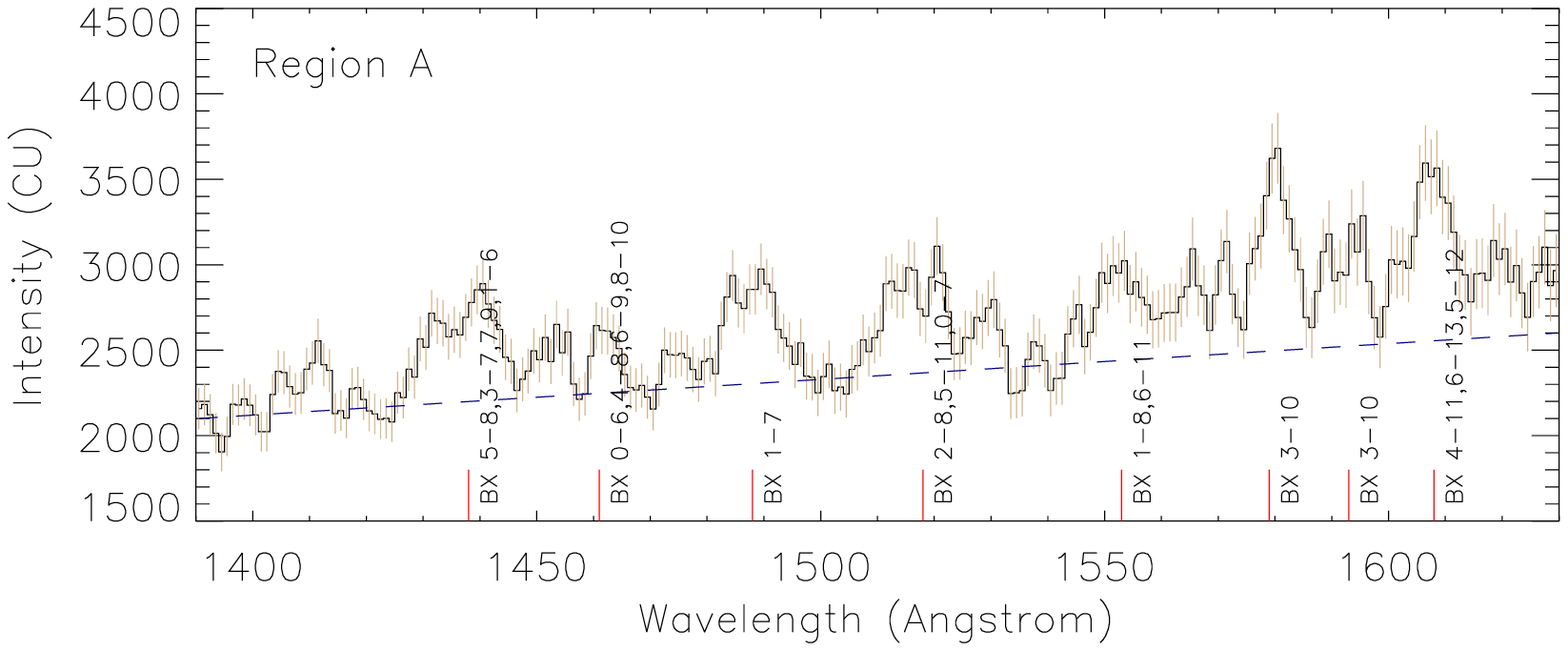} \\
  		\includegraphics[width=8cm]{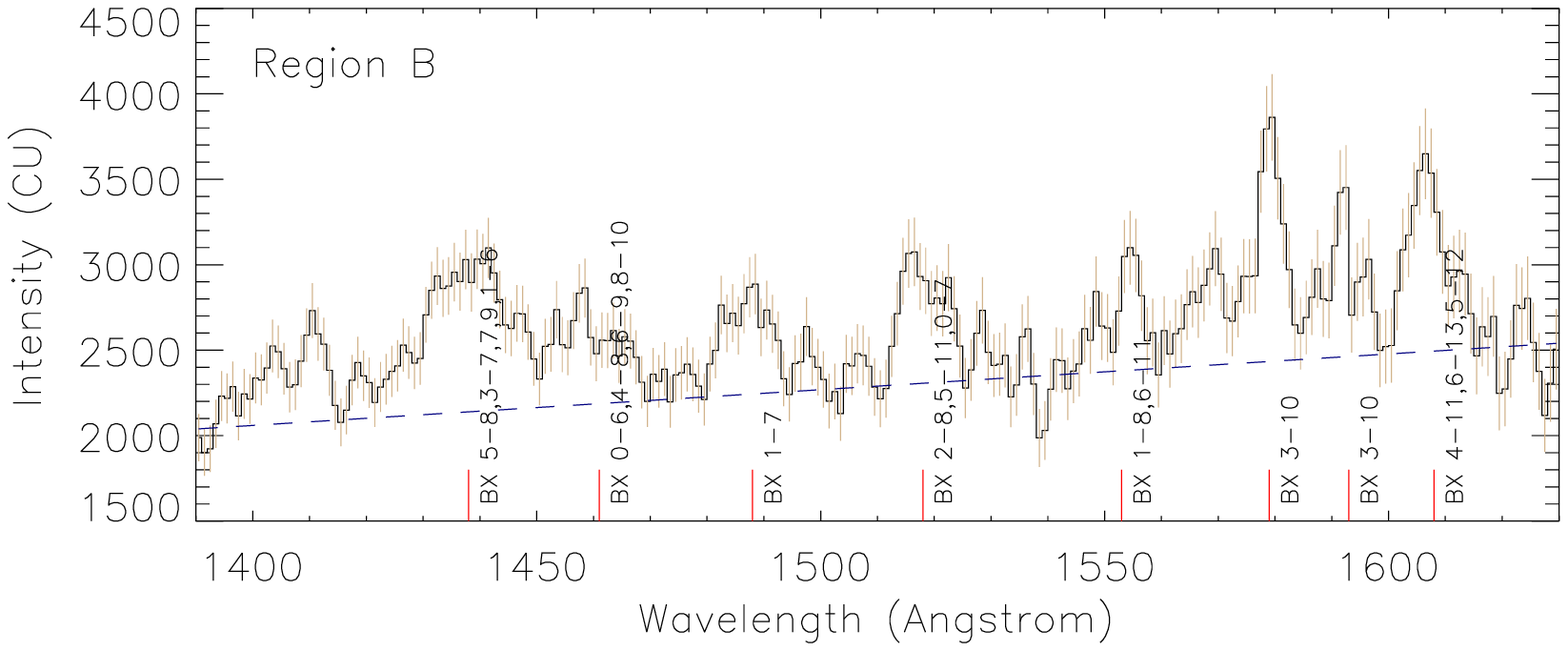} &
  		\includegraphics[width=8cm]{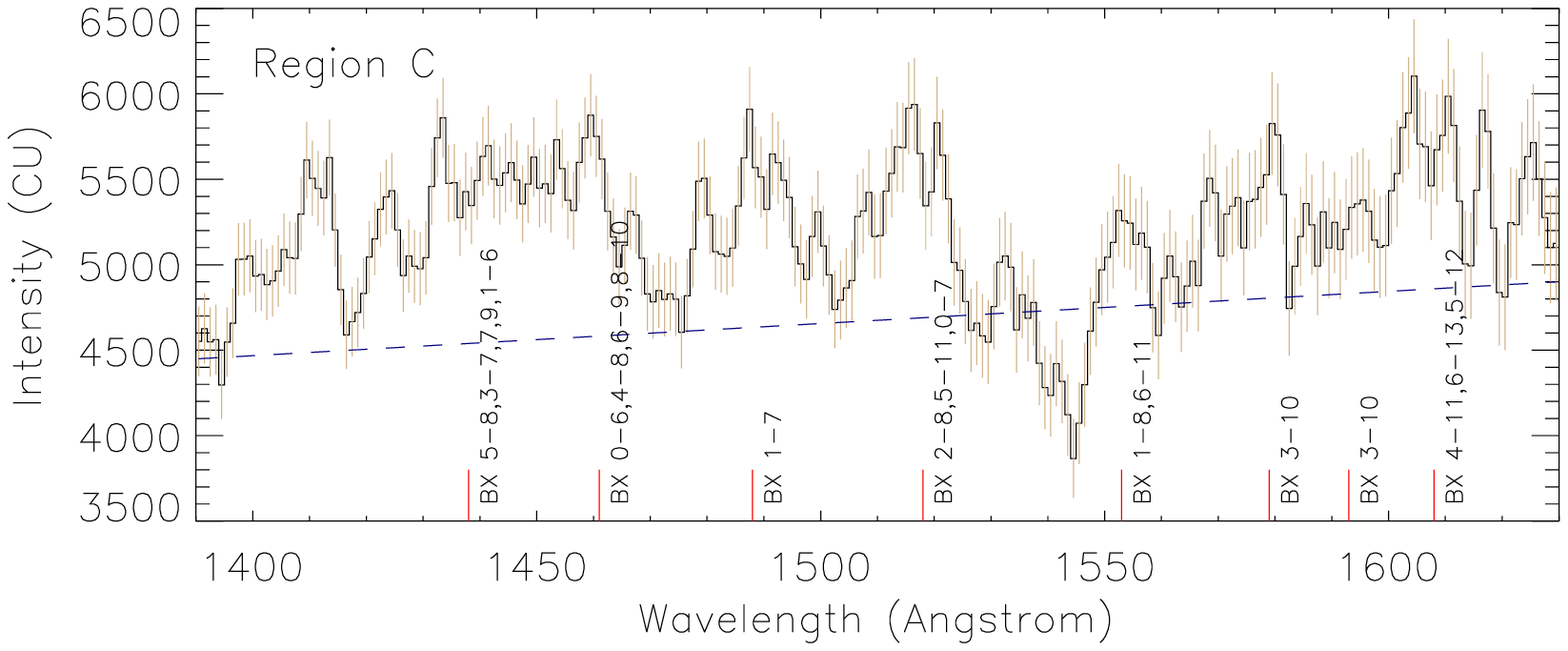}
	\end{tabular}
	\end{center}
	\caption{Spectra of the whole TPA region and subregions A, B, and C, in which the fluorescent emission lines of molecular hydrogen are indicated with vertical red bars. The dashed blue lines represent simple models for the continuum emissions.  \label{fig:spectrum}}
\end{figure*}

\begin{figure*}
	\begin{center}
	\begin{tabular}{cc}
  		\includegraphics[width=8cm]{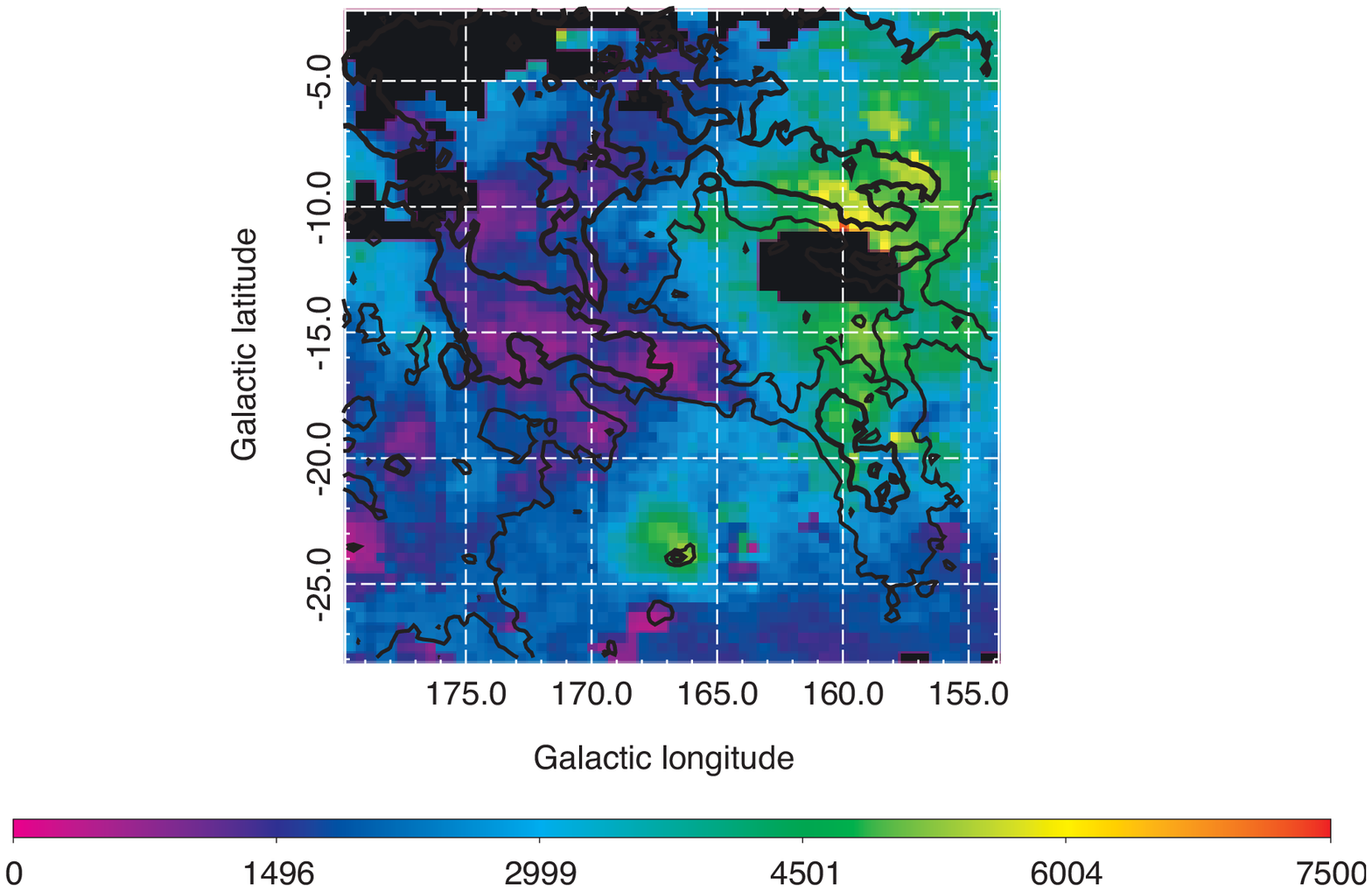} &
  		\includegraphics[width=8cm]{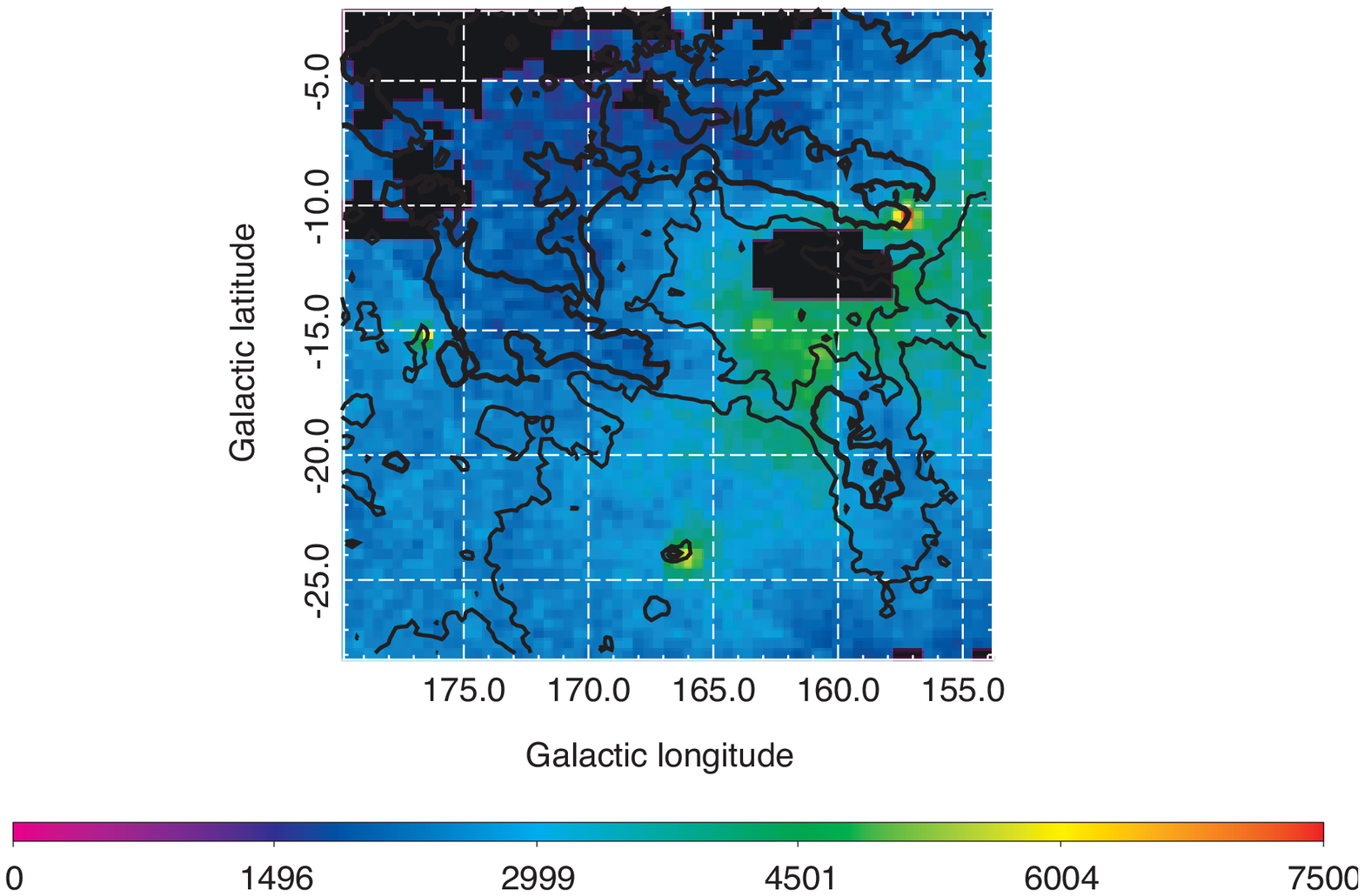}
	\end{tabular}
	\end{center}
	\caption{(a) Dust scattered FUV map constructed from the observations and (b) the Monte Carlo simulation results. The California nebula region is excluded from our dust scattering simulations as its H$\alpha$ intensity is extremely high and the associated two-photon effect is difficult to estimate.  \label{fig:simnobs}}
\end{figure*}

Our main purpose in the present study is to understand through simulations the dust scattering properties and the morphological structures in this broad region. More specifically, we seek to understand the distances and thicknesses of the TPA clouds, which includes both thick clouds and strong radiation fields of OB associations, based on the observed FUV emission features. Hence, it is necessary to identify and remove contributions other than the dust-scattered emission from the observed FUV image. In addition to dust-scattering of stellar photons, diffuse interstellar FUV emissions can have three different origins including: i) ion line emission coming from hot gas of 10$^{4.5} -$ 10$^{5}$ K; ii) H$_{2}$ fluorescent emission originating from the outer regions of molecular clouds irradiated by interstellar UV photons; and iii) two-photon effects which are mostly relevant to ionized regions. Of the other FUV sources, ion line emissions should be negligible in the present region of thick clouds. On the other hand, we expect significant H$_{2}$ fluorescent emissions in this region where both thick clouds and strong radiation fields exist. Based on the spectral observations of FIMS, Figure \ref{fig:spectrum} indeed shows that H$_{2}$ fluorescent emission is detectable throughout the whole TPA region. The contribution from the H$_{2}$ fluorescent emission lines is estimated to be $\sim$10.8 \%, $\sim$11.4 \%, and $\sim$9.5 \% of the total FUV intensity for the regions A, B, and C, respectively, and $\sim$10\% for the whole region. We have accordingly subtracted 10 \% of the total intensity from the map shown in Figure \ref{fig:multiwave}(a) to account for the contribution from H$_{2}$ fluorescent emission. Estimation of the two-photon effect can be a little tricky since H$\alpha$ emission can also originate from regions other than ionized regions, and H$\alpha$ photons may also be scattered in a manner similar to the FUV continuum photons. Furthermore, \citet{rey90} suggested a two-photon continuum emission of $\sim$60 CU for 1 Rayleigh of H$\alpha$ line in the case of a diffuse ionized gas, while \citet{seo11} estimated it to be $\sim$26 CU for 1 Rayleigh of H$\alpha$ emission of a cooling ionized gas. Hence, we have excluded the California nebula region, where H$\alpha$ intensity is especially high, in our simulation and applied the formula of the FUV continuum intensity of 60 CU for 1 Rayleigh intensity of H$\alpha$ throughout the region as a simple approximation. We will discuss later the dependence of the optical properties of dust on the models of the two-photon effects. Finally, we have subtracted uniformly an isotropic FUV emission of $\sim$300 CU \citep{bow91} although this value was obtained for the high Galactic latitude region. The resulting FUV map is shown in Figure \ref{fig:simnobs}(a).

We employed a Monte Carlo Radiative Transfer technique, in which photons were allowed to have multiple scatterings \citep{wit77,woo99,seo09,seo12}. At each scattering, after moving a mean free path distance determined by the optical depth, the photons were randomly scattered into an angle with a probability given by the Henyey-Greenstein function \citep{hen41}. We adopted the peeling-off technique to achieve simulation efficiency, i.e., regardless of the actual scattering angle determined randomly, the probability that the photon would be scattered in the direction toward the Sun was calculated for each scattering, and the corresponding fraction of the intensity was added up for the final image \citep{yus84}. We did not consider any wavelength changes upon scattering. The free parameters to be determined by the simulation were the albedo, the asymmetry factor, and the distribution of dust, while the initial photon sources were chosen to be the bright stars in the TPA region. There are, however, infinite numbers of possible choices for the three-dimensional spatial distribution of dust under the given information of integrated extinction along the lines of sight, which we adopted from the SFD dust extinction map \citep{sch98}. Therefore, we assumed, for convenience that the dust was distributed uniformly inside a slab whose thickness and distance from the Sun were to be determined. The E(B-V) values of the SFD map were converted to optical depths at 1565\AA, a representative FUV wavelength of the present study, with an assumed R$_V$ value of 3.1.

\begin{deluxetable*}{lllll} 
	\tablecolumns{10} 
	\tablewidth{0pc}
		\tablecaption{Distance to the front face and the thickness of the four selected clouds, obtained by treating each cloud individually. \label{tbl:simul}} 
		\tablehead{
  		\colhead{clouds}
	  	& \colhead{Taurus}
	  	& \colhead{Perseus}
	  	& \colhead{Auriga}
	 	& \colhead{California}}
	\startdata
	Distance to the front face (pc) & 120 $^{+20}_{-80}$ & 70 $^{+30}_{-20}$ & 140 $^{+20}_{-90}$ & 170 $^{+10}_{-10}$ \\
	Thickness (pc) & 110 $^{+20}_{-50}$ & 230 $^{+30}_{-40}$ \tablenotemark{a} & 190 $^{+70}_{-20}$ \tablenotemark{a}& 320 $^{+50}_{-30}$ \tablenotemark{a}\\
	Distance to the center (pc) & $\sim$175 & $\sim$185 & $\sim$235 & $\sim$330
	\enddata
	\tablenotetext{a}{The large thicknesses of the Perseus, California, and Auriga clouds may be due to multiple layers. Refer to Section 4 for their discussions.}
\end{deluxetable*}

The simulation box consisted of a 400 $\times$ 400 $\times$ 400 pixel, with each pixel equivalent to 1 pc. The Sun was located at the center of the front face of the simulation box. A total of 88307 stars was employed as the radiation sources, consisting of 18212 TD-1 stars and 70095 Hipparcos stars. The luminosities of the stars were calculated using the observed FUV stellar fluxes for the TD-1 stars and Castelli's stellar models for the Hipparcos stars, based on the listed spectral types, with extinction effects taken into account \citep{tho78,per97,cas03}. As a first step, we assumed a single dust slab model for the entire TPA region, which was to be improved at the second step by considering individual clouds separately, using the values of the albedo and the asymmetry factor obtained in the first step. Simulations were carried out for the following parameter ranges: the albedo values varied from 0.20 to 0.60 with 0.01 steps, the asymmetry factor varied from 0.0 to 0.80 with 0.01 steps, the distance to the front face varied from 10 pc to 390 pc with 10 pc steps, and the thickness of the slab varied from 10 pc to 390 pc with 10 pc steps. The best-fit parameters were determined by comparing the simulation results with the observed FUV map using the chi-square minimization.

\begin{figure*}
	\begin{center}
	\begin{tabular}{cc}
  		\includegraphics[width=9cm]{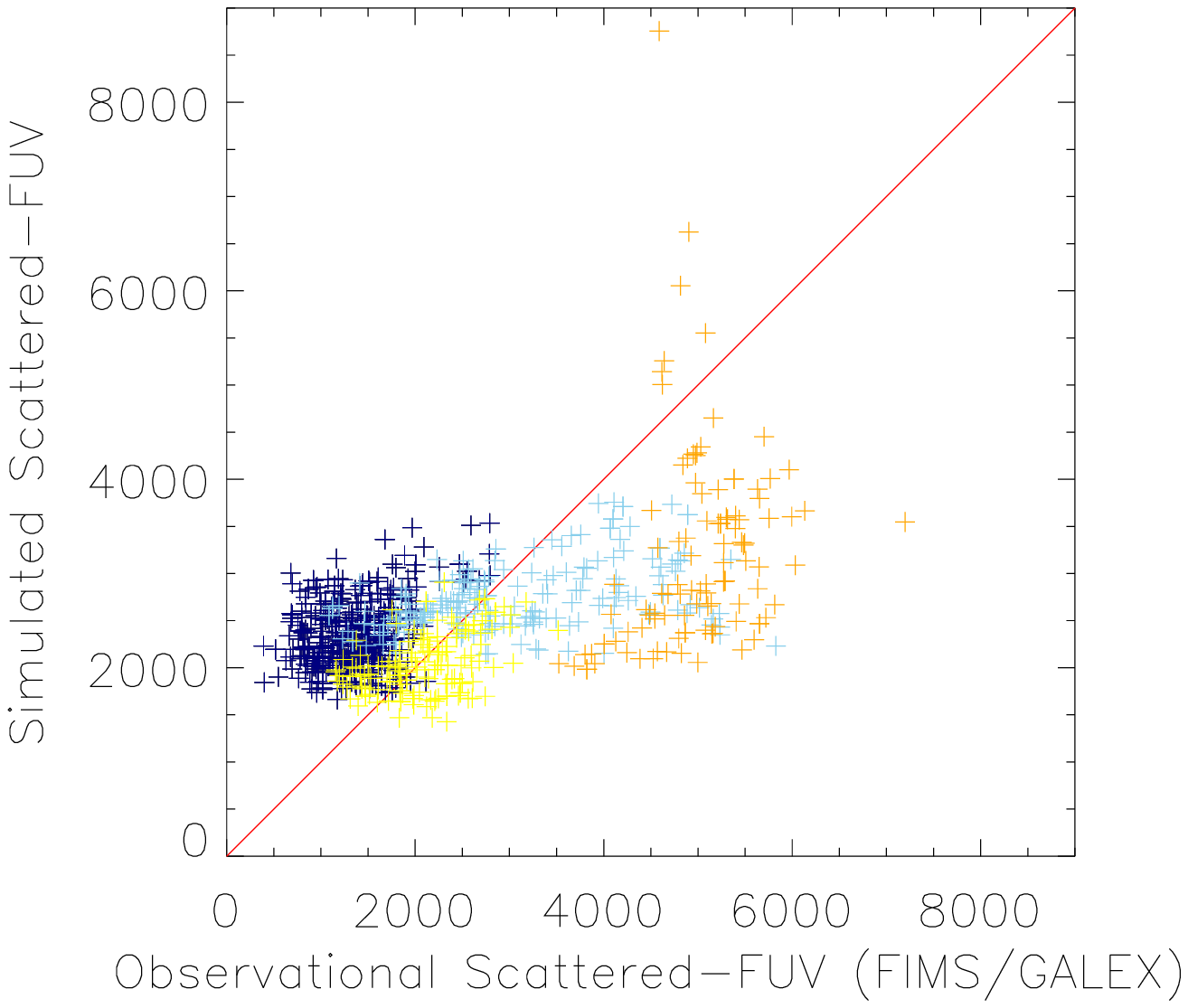} &
  		\includegraphics[width=9cm]{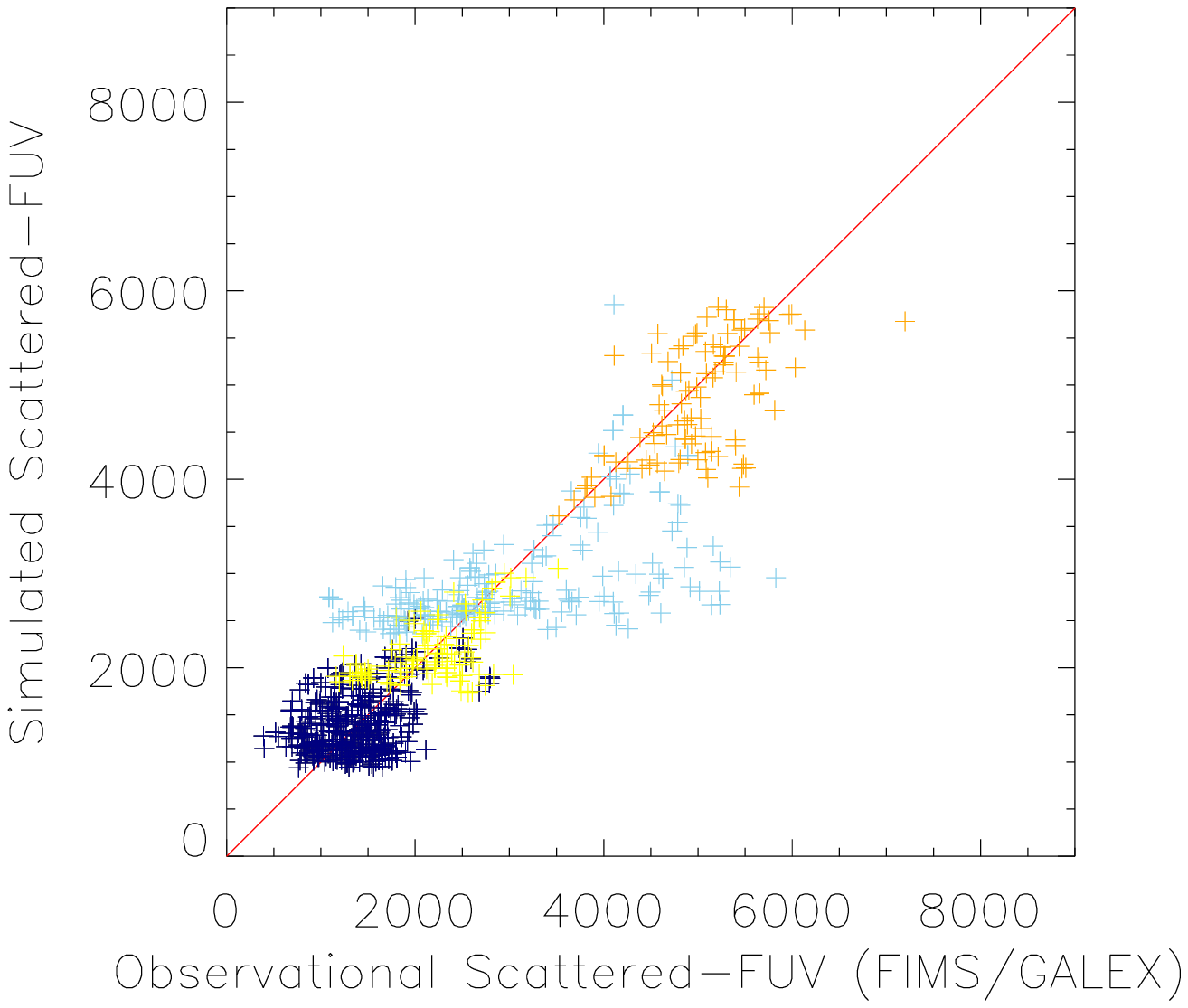}
	\end{tabular}
	\end{center}
	\caption{A scatter plot of the simulated FUV intensity against the observed FUV intensity for the regions of the Taurus, Perseus, Auriga, and California clouds. Panel a) shows the result obtained using a single dust slab model and panel b) shows the result obtained with the individual clouds treated separately. The navy, sky blue, yellow, and orange colors correspond to the regions of the Taurus, Perseus, Auriga, and California clouds, respectively.  \label{fig:simnobs_scatt}}
\end{figure*}

\begin{figure*}
	\begin{center}
	\begin{tabular}{cc}	
  		\includegraphics[width=5cm]{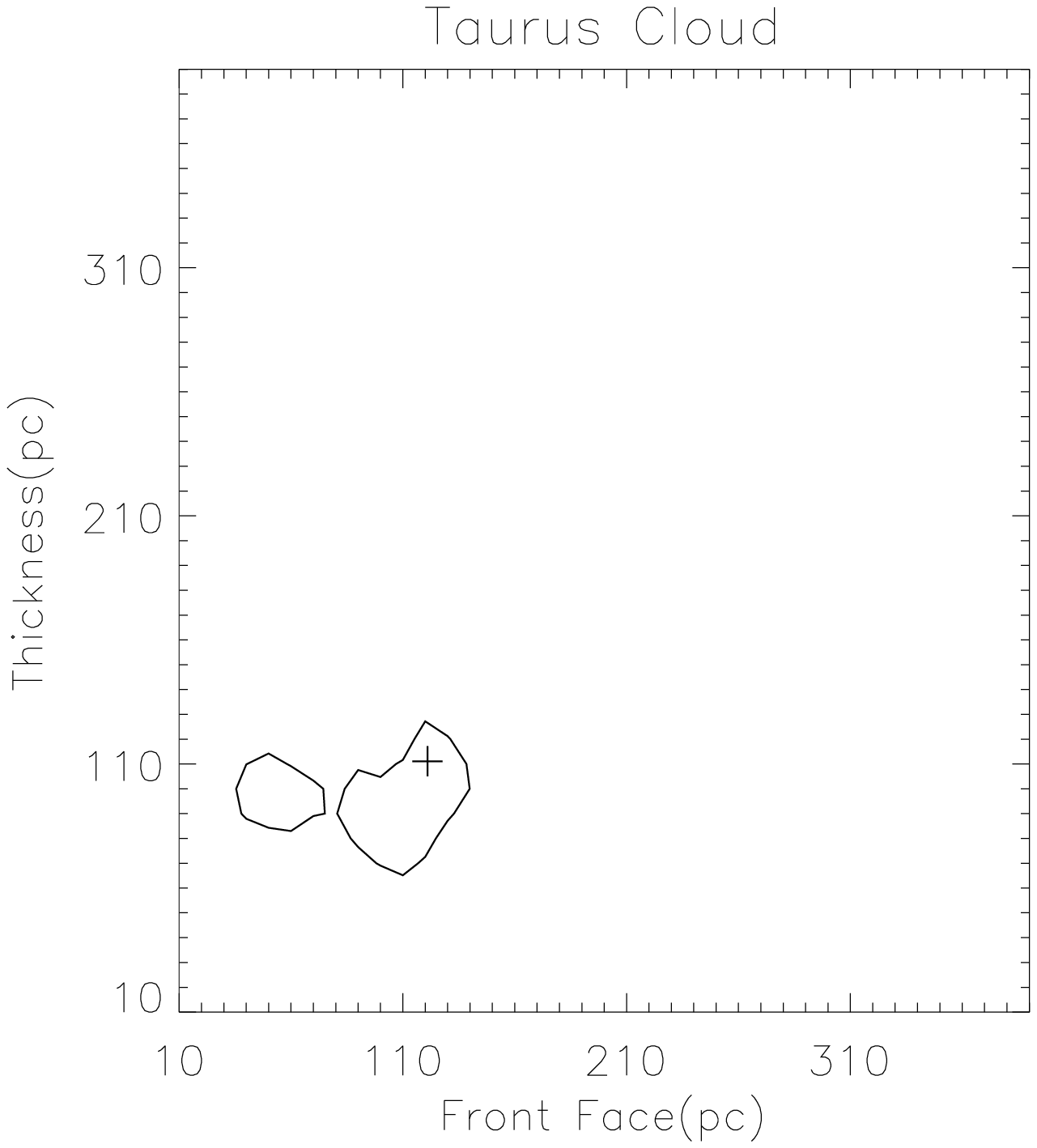} &
  		\includegraphics[width=5cm]{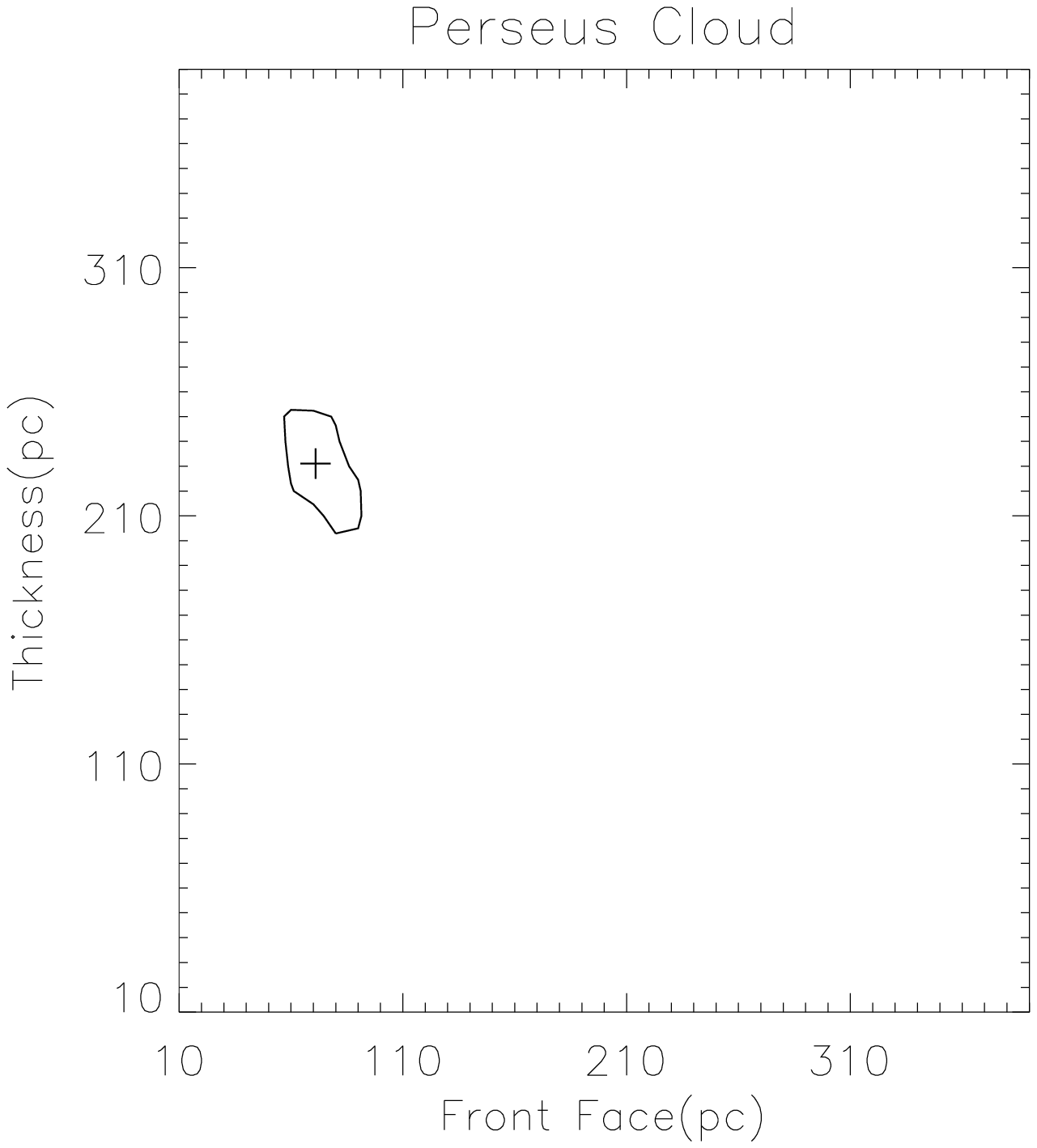} \\
  		\includegraphics[width=5cm]{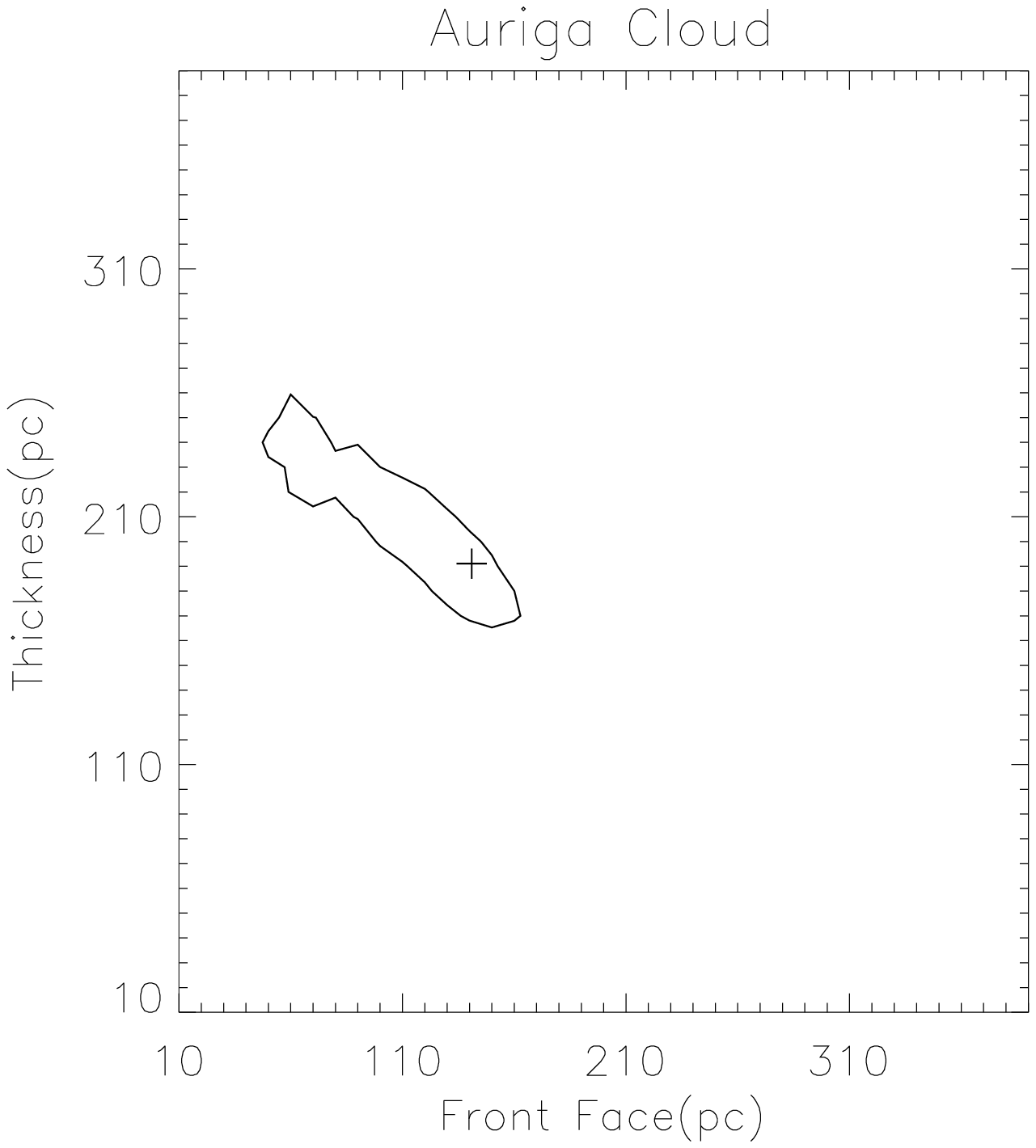} &
  		\includegraphics[width=5cm]{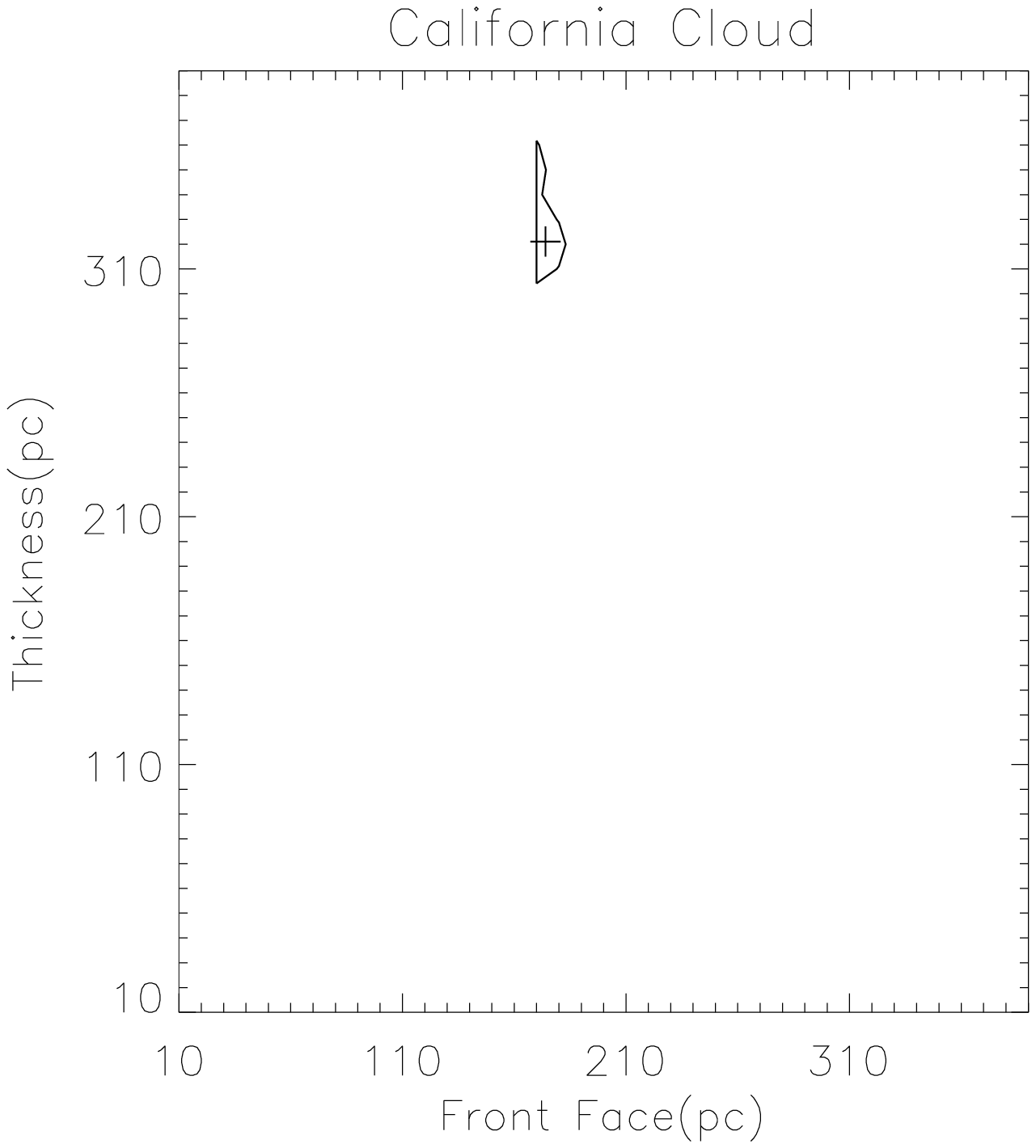}
	\end{tabular}
	\end{center}
	\caption{1-$\sigma$ confidence contours of the distance to the front face and the thickness, plotted for the four clouds identified in Figure \ref{fig:multiwave}(d). The cross symbol indicates the best-fit values. For this simulation, fixed values of the albedo (a = 0.42) and the asymmetry factor (g = 0.47) were used. \label{fig:contour}}
\end{figure*}

Figure \ref{fig:simnobs}(b) shows the simulated map with the best-fit parameters: 0.42 for the albedo, 0.47 for the asymmetry factor, 50 pc for the distance to the front face, and 220 pc for the thickness. The error ranges of the parameter values are mainly due to the statistical fluctuations of the FIMS data with low exposure time: the 1-$\sigma$ ranges of the albedo and the asymmetry factor are 0.37 to 0.47 and 0.20 to 0.58, respectively. Nevertheless, the estimated albedo of 0.42 is close to the theoretical model value of $\sim$0.40, while the asymmetry factor is rather small compared to the theoretical expectation of $\sim$0.65 \citep{wei01}. \citet{jo12} obtained 0.43 and 0.45 for the albedo and the asymmetry factor, respectively, in their simulation for the Orion-Eridanus Superbubble. The simulated map shown in Figure \ref{fig:simnobs}(b) looks similar to the observed FUV map of Figure \ref{fig:simnobs}(a), but we note that the region around the California nebula is less bright and the Taurus region is less dark in the simulated map than in the observed map. Indeed, the intensity varies less over the whole TPA region in the simulated map than in the observed map. The reason for the reduced variability could be the assumption of a single slab model for the entire region. For example, the reduction of intensity in the region around the California nebula was compensated by an increase in intensity in the Taurus region as the reduced chi-square was to be minimized with a single dust slab model. Hence, it should be meaningful for the individual clouds to be treated separately with their own distances and thicknesses. 
With the values of the albedo (a = 0.42) and the asymmetry factor (g = 0.47) determined from the single slab model, we carried out simulations separately for the individual clouds of Taurus, Perseus, Auriga, and California. For the California cloud, we enlarged the simulation box to 400 pc $\times$ 400 pc $\times$ 800 pc as the cloud was found to extend beyond 400 pc. Relative to the single slab model, the pixel-to-pixel scatter plot of the simulated FUV intensity against the observed FUV intensity shows significant improvement, as can be seen in Figure \ref{fig:simnobs_scatt}. Table \ref{tbl:simul} shows the improved distances and thicknesses of the TPA clouds, and Figure \ref{fig:contour} shows contour plots for 1-$\sigma$ error ranges. These best-fit values are in good agreement with the previous estimations based on other methods, as will be discussed in the following section. In the cases of the Taurus and Auriga clouds, the distances to the front faces seem to have rather large error ranges in the lower limit. Unfortunately, very few bright stars exist in front of these clouds that can constrain well the distances to their front faces.

\begin{figure*}
	\begin{center}
	\begin{tabular}{cc}
  		\includegraphics[width=6cm]{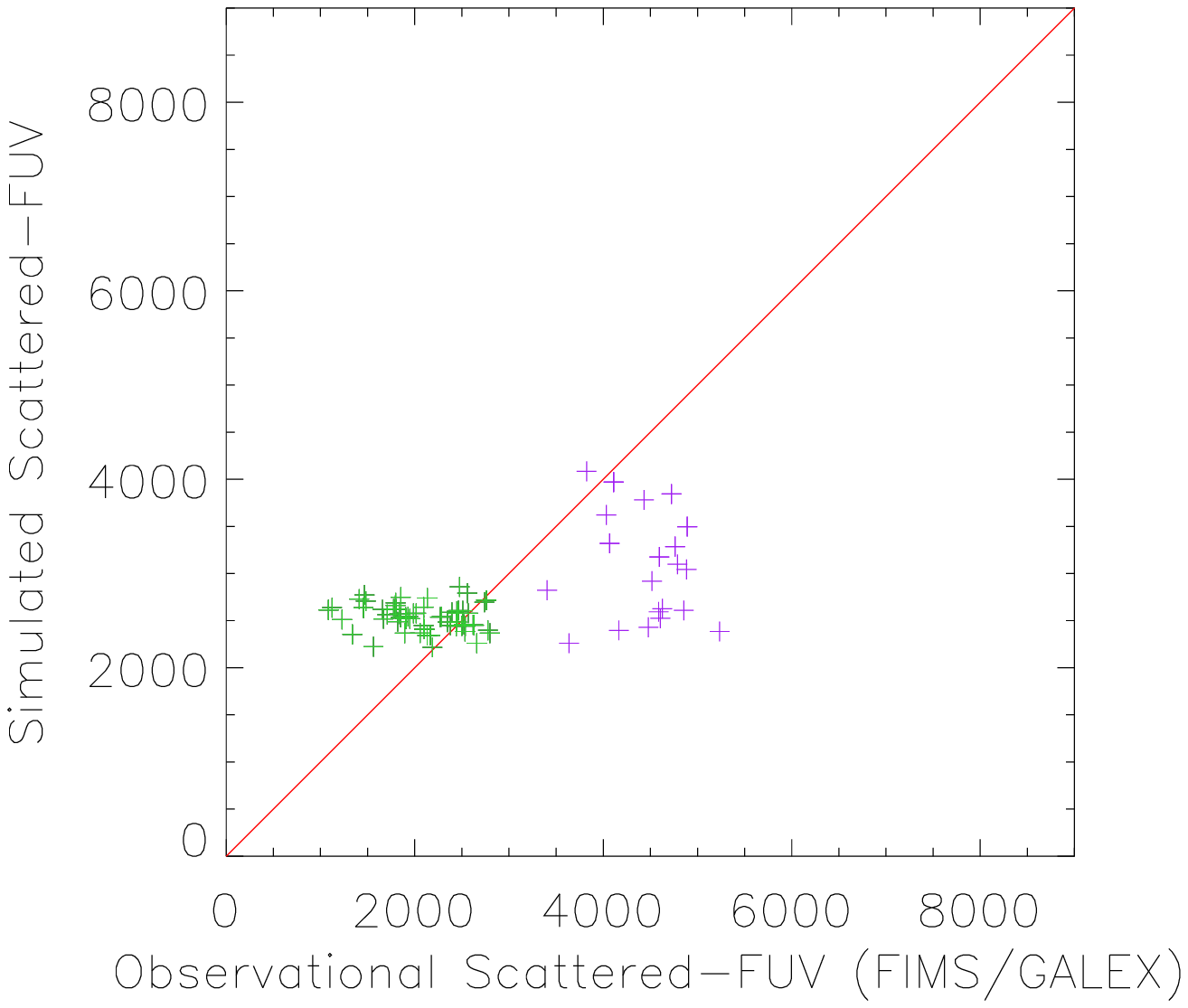} &
  		\includegraphics[width=6cm]{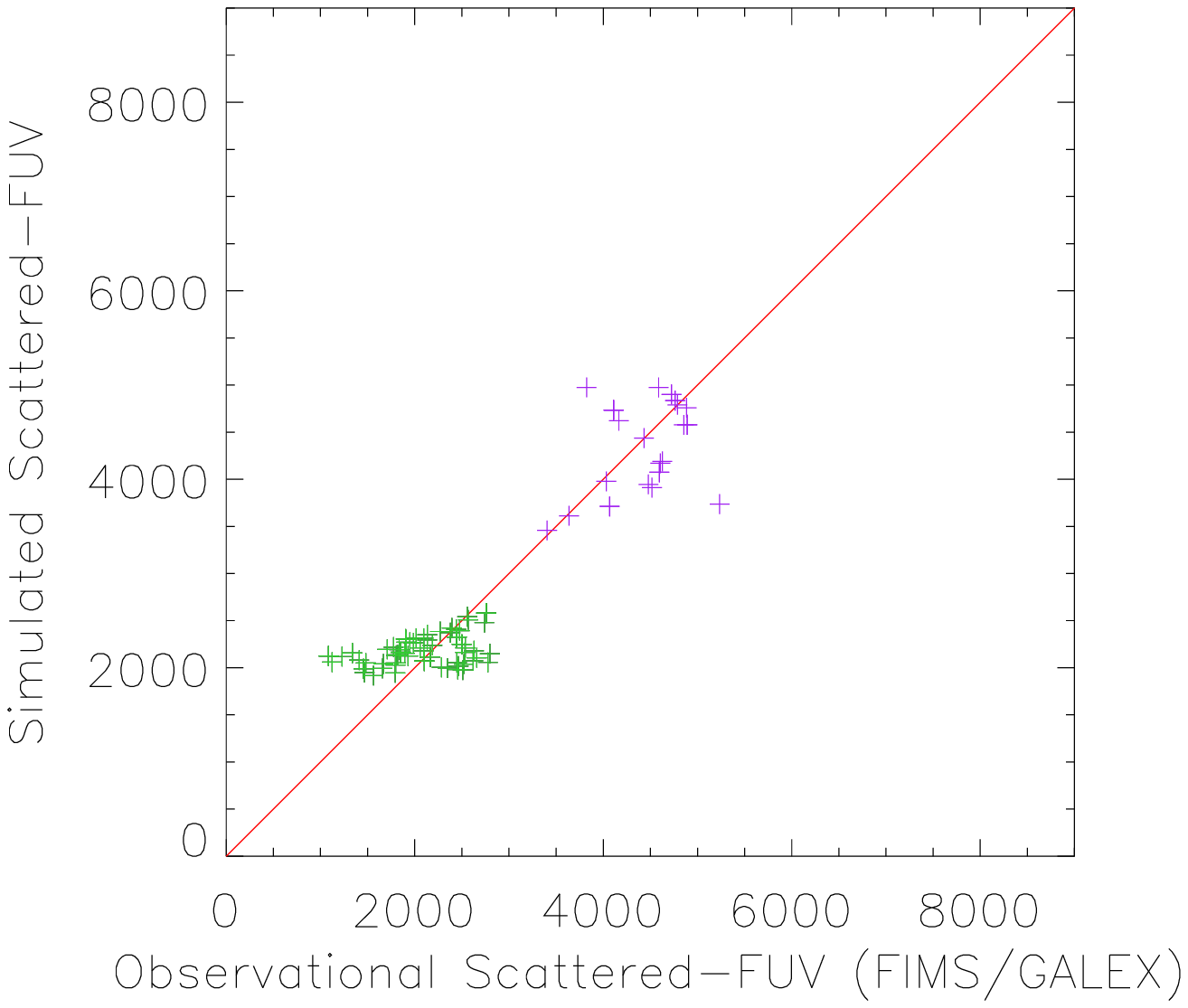}
	\end{tabular}
	\end{center}
	\caption{A scatter plot of the simulated FUV intensity against the observed FUV intensity for the two subregions A and B of the Perseus cloud, marked in Figure \ref{fig:multiwave}(d): (a) when the whole system of the Perseus cloud is treated as a single layer, and (b) when the two sub-regions are treated individually. The green and purple data points correspond to the regions of A, where NGC 1333 is located, and B, where IC 348 is located, respectively.\label{fig:scatt_subregion}}
\end{figure*}

\section{Discussion}

We have estimated the locations of the four prominent clouds in the TPA region by comparing the observed FUV images with the results of the Monte Carlo simulations of dust scattering. For the Taurus cloud, we obtained a distance to the front face and a thickness of the cloud of $120^{+20}_{-80}$ pc and $110^{+20}_{-50}$ pc, respectively. There have been many reports on the physical structure of the Taurus cloud. \citet{eli78} estimated the mean distance to the Taurus complex to be $\sim$135 pc, based on the distances to the stars associated with the Taurus cloud. \citet{ung87} suggested that the cloud spreads between 100 pc and 200 pc, based on the results of star counting, the distances of the reflection nebulae, and the color excess of the field stars. \citet{str01} also reported that extinction begins to increase sharply at 130-180 pc. The present result agrees well with all these observations.
For the Perseus cloud, the present simulation estimates a distance to the front face and a thickness of $70^{+30}_{-20}$ pc and $230^{+30}_{-40}$ pc, respectively. The scatter plot of the simulated vs. the observed FUV intensity in Figure \ref{fig:simnobs_scatt}(b) shows severely scattered data points for the Perseus cloud, indicating that the model cloud does not adequately describe its actual structure. In fact, the data points corresponding to the observation are spread over a wide range of intensity, which we believe comes from the assumption of a single layer for the cloud. Hence, we have divided the Perseus cloud into two sub-regions and fit them separately. We name Per A as the western part of the cloud and Per B as the eastern part, as shown in Figure \ref{fig:multiwave}(d), and these include the two regions NGC 1333 and IC 348, respectively. With the same fixed values of the albedo (a = 0.42) and the asymmetry factor (g = 0.47) as used in the previous simulations, we find a distance to the front face and a thickness of the Per A region of $100^{+30}_{-30}$ pc and $180^{+30}_{-50}$ pc, respectively, and a distance to the front face and a thickness of the Per B region of $\sim$320 pc and $\sim$30 pc within $^{+10}_{-10}$ pc error range, respectively. The scatter plot is now improved considerably, as shown in Figure 9. For the western portion of the Perseus cloud, \citet{cer90} obtained $\sim$170 pc and $\sim$230 pc for the distances of the two layers in this region, while \citet{hir08}, using the parallax observation of the 22 GHz H$_{2}$O maser, directly measured the distance of NGC 1333 to be $\sim$240 pc. The rather thick cloud obtained in the present simulation for Per A may reflect the multi-layer nature of the cloud suggested by \citet{cer90}. The distance to the eastern portion of the Perseus cloud containing IC 348 was estimated to be larger than 300 pc \citep{her83}. It is interesting that the thickness was estimated to be rather small for Per B in the present simulation. It was seen that the intensity of the scattered emission was very high when the cloud was placed at a distance below 300 pc. We believe the emission observed in the Per B region comes from backward scattering reflected by the cloud itself.

\begin{figure*}
	\begin{center}
  		\includegraphics[width=10cm]{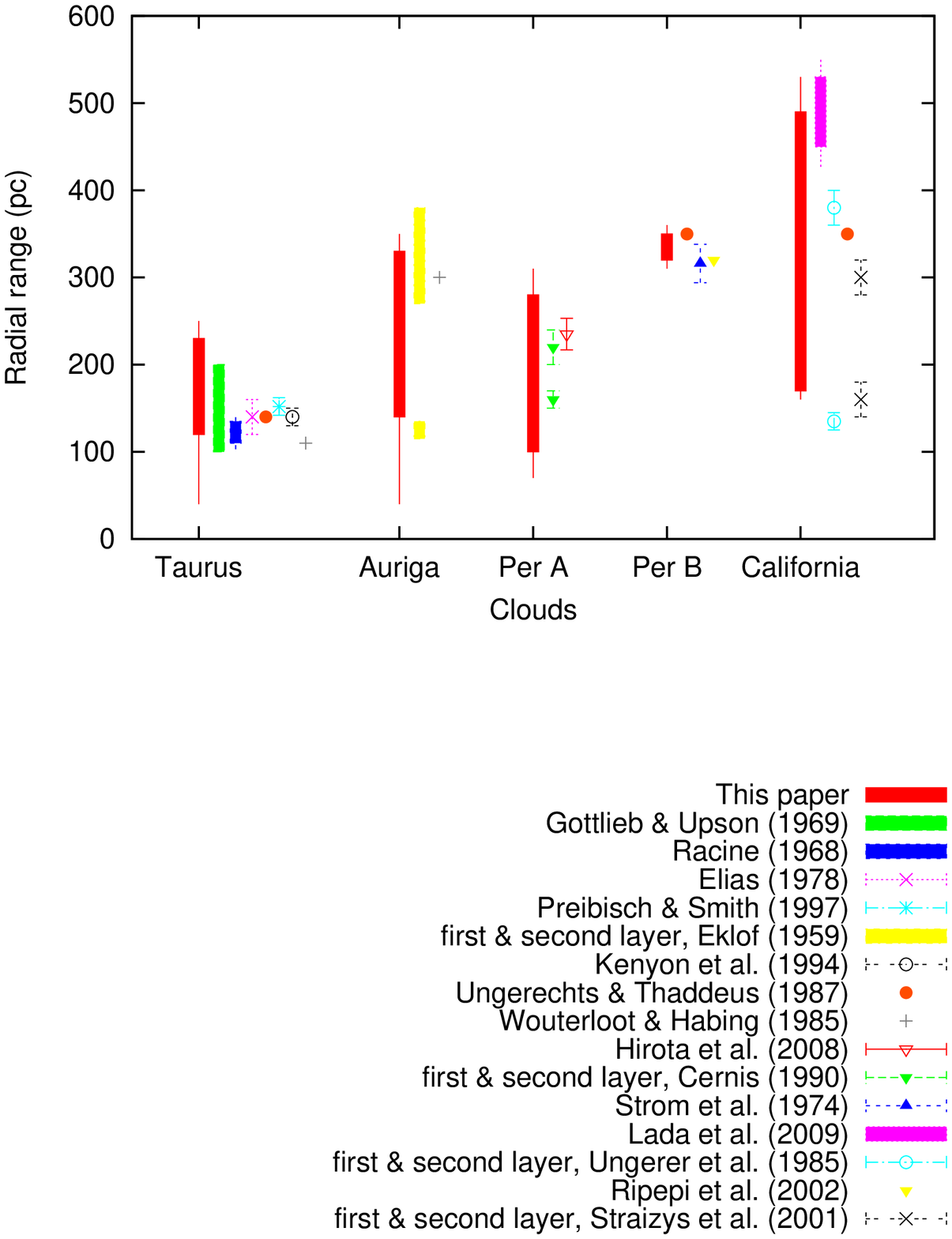}\\
	\end{center}
	\caption{Distance estimations of the four clouds based on the present Monte Carlo simulations of diffuse FUV emissions, compared with previous observations. \label{fig:summary}}
\end{figure*}

For the Auriga cloud, the distance to its front face and its thickness are estimated in the present study to be $140^{+20}_{-90}$ pc and $190^{+70}_{-20}$ pc, respectively, and for the California cloud, they are $170^{+10}_{-10}$ pc and $320^{+50}_{-30}$ pc, respectively. \citet{ekl59} suggested that the Auriga cloud has a two-layered structure near the eastern end of the California cloud, with the layers located at $\sim$115-135 pc and $\sim$270-380 pc, based on an extinction study. \citet{ung87} determined a distance to the California cloud of $\sim$350 pc while \citet{lad09} suggested that the cloud is located at $\sim$450 pc from the Sun. \citet{str01} argued that the California cloud region is two-layered, with the first layer at $\sim$160 pc and the second layer at $\sim$350 pc. With the present single-slab model, it is impossible to reproduce such multi-layer structures. Instead, we obtained single thick clouds, encompassing two layers of the clouds. We note that the estimated value of $\sim$330 pc to the center of the California cloud is larger than the distance to the Per OB2 association ($\sim$300 pc on average, \citet{bel02b}). Hence, the California cloud, being located behind the Per OB2 association which is a candidate source for the bright FUV emission seen in this region, does not block the emission from the association. Figure \ref{fig:summary} shows a summary of our distance estimations for the four clouds, along with previous observations for the corresponding clouds.

\begin{figure*}
	\begin{center}
  		\includegraphics[width=11cm]{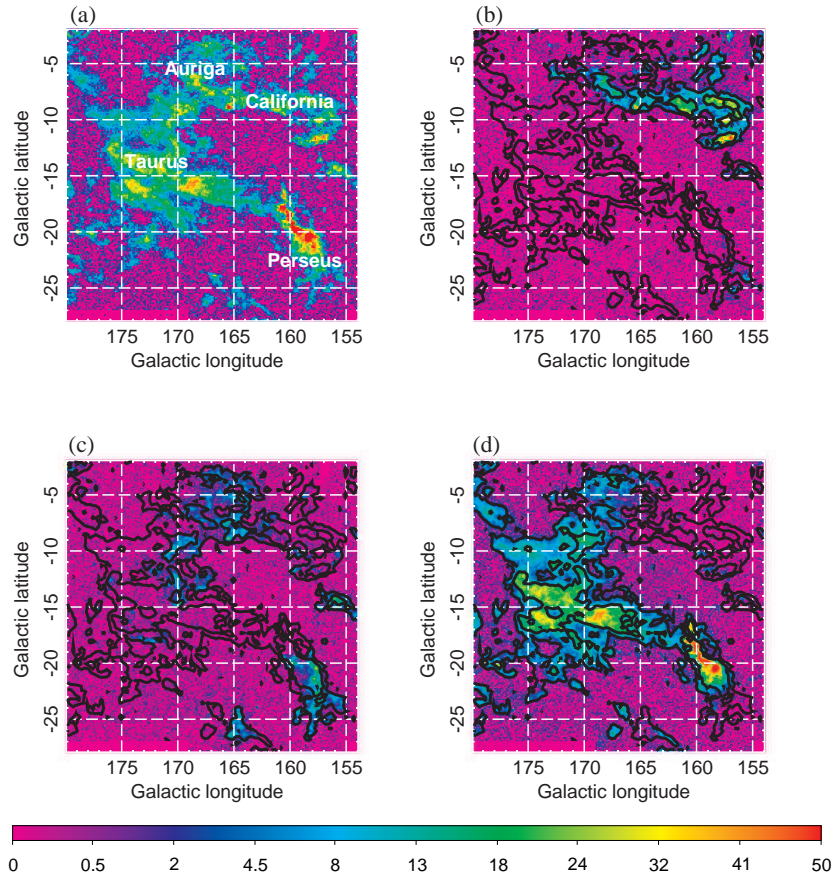}\\
	\end{center}
	\caption{Velocity-clipped CO emission maps for the ranges (a) from -21.8 km/s to +30.2 km/s, (b) from -10.7 km/s to -0.3 km/s, (c) from +0.3 km/s to +2.9 km/s, and (d) from +4.2 km/s to +10.7 km/s. \label{fig:dameco}}
\end{figure*}

We have plotted in Figure \ref{fig:dameco} the velocity-clipped CO emission maps for the ranges given in the Local Standard of Rest (LSR) frame: (a) from -21.8 km/s to +30.2 km/s, (b) from -10.7 km/s to -0.3 km/s, (c) from +0.3 km/s to +2.9 km/s, and (d) from +4.2 km/s to +10.7 km/s. These maps are reproduced from the CO sky survey data by \citet{dam01}. While the differences in the LSR velocities do not necessarily imply their relative distances, they may give insightful clues to the physical associations of the corresponding clouds. For example, the California cloud is seen in Figure \ref{fig:dameco}(b) and no other clouds are seen in that velocity range. In Figure \ref{fig:dameco}(d), however, we can see indications of the Taurus, Auriga, and Perseus clouds, but not of the California cloud. Such a grouping may indicate that the California cloud is physically not connected to other clouds, while the rest of the clouds system may be more or less associated. Furthermore, parts of the Auriga and Perseus clouds are also seen in Figure \ref{fig:dameco}(c). Hence, the fact that these clouds are seen in a wide range of the LSR velocities may be consistent with them having multi-layered structures. \citet{bal08} also noted that the large velocity gradient observed in the Perseus cloud may be due to the nature of multi-components superimposed along the line of sight.

The error ranges of the estimated optical properties are a little large, especially in the value of the asymmetry factor, where the reason was attributed earlier to the statistical fluctuations of the FIMS data with low exposure time. Nevertheless, the best fit values of the distances to the respective front face and the thicknesses of the clouds do not seem to change much when even the upper or lower limits of these error ranges are used for fitting. For example, the changes are mostly $\sim$10 pc for the distances to the front faces and $\sim$30 pc for the thicknesses of the clouds. The most sensitive parameters for fitting are instead the locations of the field stars as the photon sources relative to the clouds, not the optical properties of dust. We have also made a number of assumptions in the present study to obtain the optical properties of dust and the geometrical structures of the clouds in the TPA region. First, we used a conversion formula for the FUV continuum intensity of 60 CU for 1 Rayleigh intensity of H$\alpha$ to estimate the two-photon effects, which holds only for the case of a hot diffuse ionized gas whose temperature is $\sim$8000 K \citep{seo11}. To see the effect of the choice of the conversion factor, we ran the same Monte Carlo simulation by but instead subtracting the two-photon effect with a conversion factor of 30 CU for 1 Rayleigh of H$\alpha$ in one case and even with no subtraction in another case. The results were a = 0.43, g = 0.49 for the former case, and a = 0.44, g = 0.52 for the latter case. All these values are within the 1-sigma range of the nominal case, a = $0.42^{+0.05}_{-0.05}$ and g = $0.47^{+0.11}_{-0.27}$, implying that the choice of the conversion factor does not alter the conclusions made in the present study. Next, we subtracted 300 CU as an isotropic background of the FUV emission, which was obtained for high Galactic regions. When we subtracted 500 CU instead, only the thicknesses of the clouds changed while the distances to the respective front face remained the same. The thickness decreased by $\sim$10 \% for the Auriga and the California clouds as well as Per A, and $\sim$30 \% for the Taurus cloud region, which is darkest of all and thus most affected. The reason for the decrease in thickness is that reduction in FUV intensity was accomplished by increasing the shielding effect in the case of these clouds. On the other hand, the thickness of Per B increased by $\sim$20 pc, reflecting the nature of backward scattering in the case of Per B. Nevertheless, these new values are all within the 1-sigma range of the results obtained for the nominal case.

\section{Conclusions}
We have constructed a FUV continuum map of the TPA complex using FIMS and GALEX data. The morphological features seen in the complex are well understood in terms of dust scattering, as demonstrated by Monte Carlo simulations. The following are the main findings of the present study:

\begin{enumerate}
\item The diffuse FUV emission seen in the complex originates mostly from scattering of stellar photons by dust grains.
\item The FUV intensity of ~1000 CU is observed at a very high extinction level, which we regard as diffuse background and attribute to scattered FUV photons located in the foreground to the thick clouds.
\item Molecular hydrogen fluorescent emission constitutes $\sim$10\% of the total FUV intensity throughout the region.
\item We have derived the following scattering parameters for this region based on the Monte Carlo Radiative Transfer (MCRT) simulations: albedo (a) = $0.42^{+0.05}_{-0.05}$ and asymmetry factor (g) = $0.47^{+0.11}_{-0.27}$. These values agree well with those obtained previously for the Orion-Eridanus Superbubble region as well as theoretical estimations.
\item We have estimated the distances to the four prominent clouds in this complex, which are in good agreement with those estimated observationally using other methods. The thickness of each cloud was also reasonably determined when it has a simple structure such as the Taurus cloud. The present single slab model, however, was not able to reproduce multi-layered clouds, but gives rather thick clouds, instead. 
\item The geometrical structures of the clouds are less sensitive to the exact values of the albedo and the asymmetry factor. Instead, the locations of the field stars relative to the clouds are the main factors that constrain the distance and thickness of the clouds.
\end{enumerate}

\acknowledgments
FIMS/SPEAR is a joint project of KAIST and KASI (Korea) and UC Berkeley (USA), funded by the Korea MOST and NASA grant NAG5-5355. This research was supported by Basic Science Research Program (2010-0023909) and National Space Laboratory Program (2008-2003226) through the National Research Foundation of Korea (NRF) funded by the Ministry of Education, Science and Technology.

\clearpage

\end{document}